%% file: paper.tex
\documentclass[a4paper]{aa}
\usepackage{amsmath}
\usepackage{epsfig}
\usepackage{graphicx}

\newcommand{\pd}[1]{\frac{\partial #1}{\partial t}}
\newcommand{\rot}[1]{\nabla \times #1}
\newcommand{\cp}[2]{#1 \times #2}
\newcommand{\grad}[1]{\nabla #1}
\newcommand{\dive}[1]{\nabla \cdot #1}
\newcommand{\vf}[1]{{\bf{#1}}}
\newcommand{\vfg}[1]{\boldsymbol{#1}}
\newcommand{\ten}[1]{\boldsymbol{\mathsf{#1}}}
\newcommand{\omt}{\widetilde{\omega}}
\newcommand{\om}[1]{\omega_{\mathrm{#1}}}
\newcommand{\Om}[1]{\Omega_{\mathrm{#1}}}
\newcommand{\kt}{k_{\mathrm{0}}}

\begin{document}

\authorrunning{J.W.S. Blokland et al.}
\title{Magneto-rotational overstability in accretion disks}
\author{J. W. S. Blokland\inst{1,2} \and E. van der Swaluw\inst{1} \and R. Keppens\inst{1,2} \and J. P. Goedbloed\inst{1,2}}
\offprints{J.W.S. Blokland,\email{blokland@rijnh.nl}}
\institute{FOM-Institute for Plasma Physics Rijnhuizen, P.O. Box 1207, 3430 BE Nieuwegein, The Netherlands
           \and Association EURATOM-FOM, Trilateral Euregio Cluster} 
\date{Recieved ;accepted}

\abstract{We present analytical and numerical studies of magnetorotational
instabilities occuring in magnetized accretion disks. These calculations are 
performed for general radially stratified disks in the cylindrical limit. We elaborate 
on earlier analytical results and confirm and expand them with numerical computations
of unstable eigenmodes of the full set of linearised compressible MHD equations. We
compare these solutions with those found from approximate local dispersion
equations from WKB analysis.

In particular, we investigate the influence of a nonvanishing toroidal magnetic field 
component on the growth rate and oscillation frequency of magnetorotational instabilities in
Keplerian disks. These calculations are performed for a constant axial
magnetic field strength. We find the persistence of these instabilities in 
accretion disks close to equipartition. Our calculations show that these
eigenmodes become overstable (complex eigenvalue), due to the presence of a toroidal 
magnetic field component, while their growth rate reduces slightly.

Furthermore, we demonstrate the presence of magneto-rotational overstabilities in 
weakly magnetized sub-Keplerian rotating disks. We show that the growth rate scales 
with the rotation frequency of the disk. These eigenmodes also have a nonzero oscillation 
frequency, due to the presence of the dominant toroidal magnetic field component.
The overstable character of the MRI increases as the rotation frequency of the disk decreases.

}
\titlerunning{Magneto-rotational overstability in accretion disks}
\maketitle
   
\section{Introduction}
Accretion disks are a rather common phenomenon in astrophysics.
They vary in size from a few up to a hundred astronomical units (AU) in young 
stellar objects (YSO) to approximately a hundred parsecs (pc) in the centers 
of active galactic nuclei (AGN). The central accreting object is a protostar 
in YSOs, a white dwarf, neutron star or a black hole in a binary, and a massive 
black hole in the center of an AGN. Part of the gravitational energy released 
during the accretion process is radiated away over a wide range of frequencies. 
Multi-wavelength studies obtained from modern observational facilities have 
significantly increased our knowledge about the nature of accretion disks and 
their associated central objects.

On the other hand, insights into the nature of the accretion process itself
have largely been obtained by theoretical and computational studies.
In all the above mentioned astrophysical objects, one needs outward
transport of angular momentum of the accreting material in order to sustain an
accretion disk around the central object. This angular momentum
transport can be sustained by a turbulent viscosity mechanism operating inside
the disk material itself, where this mechanism excerts a torque on the 
accretion disk (Shakura \& Sunyaev \cite{SS}). This turbulent viscosity, in turn,  
can originate from the development of fluid or magnetofluid instabilities occuring 
in the accretion disk. In the early 1990s it was realised by Balbus \& Hawley (\cite{HB}) 
that magneto-rotational instability (MRI) could provide the physical basis for 
this angular momentum transport in accretion disks. This instability was already known in
the literature (Velikhov \cite{V} and Chandrasekhar \cite{Chandra}), but had not been 
applied in the context of accretion disks.

In more recent years, global magnetohydrodynamical (MHD) simulations of accretion
disks have been performed, where much of the dynamics is interpreted as
a direct consequence of the presence of the
magneto-rotational instability (see for example Hawley, Balbus \& Stone \cite{HBS}). 
Other papers rather claim that both convective and magneto-rotational instabilities 
can play a dominant role in the transport of angular momentum (see for example Igumenshchev,
Narayan \& Abramowicz \cite{INA}). These types of disks are referred to as
convection-dominated accretion flows (CDAF). Less attention has
been paid in recent years to the spectral analysis of instabilities occurring
in accretion disks (see however Christodoulou, Contopoulos \& Kazanas
\cite{CCK}). 

The aim of this paper is to present a detailed 
linear analysis of magnetorotational instabilities present in a variety of global 
accretion disk configurations. We will consider a sample of accretion disk configurations 
that vary from sub-Keplerian (thick) to Keplerian (thin) rotating disks. In the case of 
Keplerian rotating disks, we will consider both weakly magnetized disks and disks that are 
close to equipartition. In particular, we investigate the influence of a nonvanishing toroidal 
magnetic field  component on the growth rate and oscillation frequency of magnetorotational 
instabilities in Keplerian disks. To our knowledge, this has not yet been done.

In this paper, we will limit ourselves to \textit{axisymmetric} instabilities and
the configurations are all taken in the cylindrical limit. The calculations
continue the work presented by Keppens, Casse \& Goedbloed (\cite{KCG}), advocating the need
for a more elaborate magnetohydrodynamic spectroscopic analysis of all waves and instabilities
in magnetized disks. We will use a semi-analytical approach, as well as numerical solutions
to the full set of linearised, compressible MHD equations obtained with the 
code LEDAFLOW (Nijboer et al. \cite{NHPG}).

We will first show that magneto-rotational instabilities are present in both sub-Keplerian 
and Keplerian rotating disks and compare the growth rates obtained for
these models. We find the presence of MRI {\it even} for cases where the
disk is close to equipartition (i.e. $\beta\sim 1$), however the strength of the toroidal
magnetic field should then be much larger than the axial magnetic field strength.

This paper is organised as follows: in section 2, we recall the essential elements from
spectral theory of MHD waves and instabilites. In section 3, we present the model
and the limitations of the accretion disk configuration we use. 
In section 4, we explain the numerical strategies. In section 5, we discuss the 
magnetorotational overstabilities which occur in our models and finally, in section 6,
we summarise and present our conclusions.

\section{Spectral theory}
We make use of the ideal MHD approximation to model an accretion disk in the presence of a magnetic field.
This assumes that the disk matter consists of a plasma that is sufficiently ionized and that one
treats the dynamics on a length scale such that the one-fluid approximation is valid. 
The ideal MHD equations are
\begin{align}
  \label{eq:momentum}
  \rho \pd{\vf{v}} & = -\rho\vf{v}\cdot\grad{\vf{v}} - \grad{p} + \cp{\vf{j}}{\vf{B}} - \rho\grad{\Phi}, \\
  \label{eq:energy}
  \pd{p}           & = -\vf{v}\cdot\grad{p} - \gamma p\dive{\vf{v}}, \\
  \label{eq:faraday}
  \pd{\vf{B}}      & = \rot{(\cp{\vf{v}}{\vf{B}})}, \\
  \label{eq:density}
  \pd{\rho}        & = -\dive{(\rho\vf{v})},
\end{align}
where the variables, $\rho$, $p$, $\vf{v}$, $\vf{B}$, $\Phi$, $\vf{j} = \rot{\vf{B}}$ and $\gamma$
are the density, pressure, velocity, magnetic field, gravitional potential, electric current and
ratio of the specific heats, respectively. Furthermore, from Maxwell's theory, the equation~$\dive{\vf{B}} = 0$ 
must also be satisfied. We have assumed a dimensionalization where the permeability of 
vacuum~$\mu_{0}=1$. 
\mbox{Equations \eqref{eq:momentum}-\eqref{eq:density}} are the momentum, entropy, induction 
and mass conservation equation, respectively.

\subsection{Frieman-Rotenberg formalism}
To investigate the stability properties of accretion disks, we linearize the ideal MHD 
equations~\eqref{eq:momentum}--\eqref{eq:density}. 
The linearisation is done by assuming a time-independent equilibrium. This assumption is justified
at time scales which are much shorter than the dynamical time scale of the disk equilibrium. 
When linearizing, we write all variables of the ideal MHD equations in the following way:
\begin{equation}
  \label{eq:lineasation}
  f = f_{0}(\vf{r}) + f_{1}(\vf{r},t), \qquad \text{and} \ |f_{1}| \ll f_{0}.
\end{equation}
Here, $f_{0}$ is the equilibrium quantity while $f_{1}$ is the time-dependent fluctuation about the equilibrium
quantity. The resulting equations can be rewritten in terms of the Lagrangian displacement 
field,~$\vfg{\xi}(\vf{r},t)$, which is related to the perturbed velocity,~$\vf{v}_{1}$, in the following way
\begin{equation}
  \label{eq:diplacement}
  \vf{v}_{1} = \left( \pd{} + \vf{v}_{0}\cdot\grad{} \right)\vfg{\xi} - \vfg{\xi}\cdot\grad{\vf{v}_{0}},
\end{equation}
where~$\vf{v}_{0}$ represents the equilibrium flow velocity. By introducing the Lagrangian displacement field, 
there is no longer confusion between equilibrium and perturbed quantities when we suppress the subscript 0 on all 
equilibrium quantities. The govering equation for the displacement field, called the Frieman-Rotenberg 
equation~(Frieman and Rotenberg \cite{FR}), is
\begin{equation}
  \label{eq:fr}
  \rho \frac{\partial^{2}\vfg{\xi}}{\partial t^{2}} + 2\rho\vf{v}\cdot\grad{\pd{\vfg{\xi}}} - \ten{F}(\vfg{\xi}) = 0,
\end{equation}
where~$\ten{F}(\vfg{\xi})$ represents the force operator, defined by
\begin{equation}
  \label{eq:forceoperator}
  \begin{aligned}
  \ten{F}(\vfg{\xi}) & \equiv -\grad{\Pi} + \vf{B}\cdot\grad{\vf{Q}} + \vf{Q}\cdot\grad{\vf{B}} 
                             - \grad{\Phi}\dive{(\rho\vfg{\xi})} \\
		    & \quad
			     + \ \dive{[ \rho\vfg{\xi}\vf{v}\cdot\grad{\vf{v}} 
  				         - \rho\vf{v}\vf{v}\cdot\grad{\vfg{\xi}} ]},
  \end{aligned}
\end{equation}
with the Eulerian perturbation of the total pressure,
\begin{equation}
  \label{eq:epressure}
  \Pi \equiv -\gamma p\dive{\vfg{\xi}} - \vfg{\xi}\cdot\grad{p} + \vf{B}\cdot\vf{Q},
\end{equation}
and the Eulerian perturbation of the magnetic field,
\begin{equation}
  \label{eq:emagnetic}
  \vf{Q} \equiv \rot{(\cp{\vfg{\xi}}{\vf{B}})}.
\end{equation}

By assuming an exponential time-dependence~$\mathrm{e}^{-\mathrm{i}\om{}t}$ for the displacement field, we 
can distinguish two cases. In the case of no equilibrium flow, the force operator~$\ten{F}(\vfg{\xi})$ 
is self-adjoint, meaning that its eigenvalues $\om{}$ are purely real or imaginary.  This results in stable, 
damped and unstable modes. When there is flow, the force operator is no longer self-adjoint, 
meaning that its eigenvalues~$\om{}$ are complex in general. This introduces the possiblity of damped stable
waves as well as overstable modes.

\subsection{System of first order differential equations}
The Frieman-Rotenberg equation~\eqref{eq:fr} will be applied in the case of a one-dimensional 
cylindrical plasma equilibrium. For this kind of equilibrium, the 
MHD \mbox{equations \eqref{eq:momentum}--\eqref{eq:density}} reduce to the radial force balance equation,
\begin{equation}
  \label{eq:forcebalance}
  \left( p + \tfrac{1}{2}B^{2} \right)' + \frac{B_{\theta}^{2}}{r} = \frac{\rho v_{\theta}^{2}}{r} - \rho g,
\end{equation}
where the prime indicates the derivative with respect to $r$. The symbols $B$, $B_{\theta}$, and $v_{\theta}$ are
the total magnetic field, toroidal magnetic field, and toroidal velocity, respectively.
Furthermore, $g$ represents the gravitational acceleration at the distance $r$,
\begin{equation}
  \label{eq:gravity}
   g = \frac{GM_{*}}{r^{2}},
\end{equation}
with~$G$ the gravitational constant and~$M_{*}$ the mass of the central object.

For the three-dimensional perturbations, we choose Fourier mode solutions
of the form
\begin{equation}
  \label{eq:fouriermodes}
  \vfg{\xi}(r,\theta,z,t) = \left(
                            \begin{array}{c}
			      \xi_{r,mk}(r)      \\
			      \xi_{\theta,mk}(r) \\ 
			      \xi_{z,mk}(r)
			    \end{array}
			    \right)
                            \mathrm{e}^{\mathrm{i}(m\theta + kz - \omega t)},
\end{equation}
where~$m$ and $k$ are the toroidal and axial wavenumber, respectively.
This choice can be made because of the symmetry of the equilibrium. 
Also by exploiting a projection based on the magnetic field lines, the Frieman-Rotenberg
equation~\eqref{eq:fr} can be reduced to a system of first order differential equations, which reads
\begin{equation}
  \label{eq:fod}
  \frac{AS}{r}
  \begin{pmatrix}
    \chi \\ \Pi
  \end{pmatrix}'
   + \left(
  \begin{array}{rr}
    C &  D \\
    E & -C
  \end{array}
  \right)
  \begin{pmatrix}
    \chi \\ \Pi
  \end{pmatrix}
  = 0,
\end{equation}
where
\begin{align}
  \chi & \equiv r\xi_{r}, \\
  A    & \equiv \rho\omt^{2} - F^{2}, \\
  S    & \equiv (\gamma p + B^{2})\rho\omt^{2} - \gamma p F^{2}, \\
  D    & \equiv \rho^{2}\omt^{4} - \kt^{2} S, \\
  C    & \equiv -\frac{\rho\omt^{2}}{r} \Biggl\{ \frac{\rho}{r} 
                \left[ (B_{\theta}^{2}-\rho v_{\theta}^{2})\omt^{2} + (B_{\theta}\omt + Fv_{\theta})^{2} \right] \\
       & \quad \qquad \qquad
		+ \rho gA \Biggr\} + 2 \frac{m}{r}\frac{S}{r^{2}} (FB_{\theta} + \rho\omt v_{\theta}),\nonumber \\
\intertext{and}
  E    & \equiv -\frac{AS}{r^{2}} 
                \left[ A + r\left( \frac{B_{\theta}^{2} - \rho v_{\theta}^{2}}{r^{2}} \right)' + \rho' g \right] \\
       & \quad
		+ 4\frac{S}{r^{4}} (B_{\theta}F + \rho\omt v_{\theta})^{2} \nonumber \\
       & \quad
		- \frac{1}{r^{2}} \Biggl\{ \frac{\rho}{r} 
                \left[ (B_{\theta}^{2}-\rho v_{\theta}^{2})\omt^{2} + (B_{\theta}\omt + Fv_{\theta}) \right] 
		+ \rho gA \Biggr\}^{2}. \nonumber
\end{align}
In these expressions we use the following definitions: $\vf{\kt} = \left[ 0,m/r,k \right]$,  $\kt=|\vf{\kt}|$, 
the Doppler shifted \mbox{frequency $\omt = \omega - \vf{\kt}\cdot\vf{v}$}, 
\mbox{and $F = \vf{\kt}\cdot\vf{B}$}. Notice that we have suppressed the subscripts~$m$ and~$k$.
The \mbox{system \eqref{eq:fod}} has been derived by 
\mbox{Hameiri (\cite{Ha})} and \mbox{Bondeson et al. (\cite{BIB})} for a cylindrical plasma with 
flow but without gravity. Recently, this system was obtained by \mbox{Keppens et al. (\cite{KCG})}.

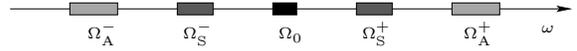
\begin{figure}[h]
  \centering
  \input{spectrum.pstex_t}
  \caption{Continuous parts of the MHD spectrum for a weakly inhomogeneous equilibrium.}
  \label{fig:spectrum}
\end{figure}

The differential equations~\eqref{eq:fod} become singular when $A = 0$ or $S = 0$, which results in the MHD continua. When $A=0$, the 
eigenfrequency~$\omega$ is equal to the local Doppler shifted Alfv\'en continuum frequency,
\begin{equation}
  \label{eq:doppleralfven}
  \Om{A}^{\pm} \equiv \vf{\kt}\cdot\vf{v} \pm \om{A},
\end{equation}
where $\om{A} \equiv F / \sqrt{\rho}$ is the local Alfv\'en continuum frequency for static equilibrium.
If $S=0$, the eigenfrequency $\omega$ is equal to the local Doppler shifted
slow continuum frequency:
\begin{equation}
  \label{eq:dopplerslow}
  \Om{S}^{\pm} \equiv \vf{\kt}\cdot\vf{v} \pm \om{S},
\end{equation}
where $\om{S} \equiv \sqrt{\gamma p / (\gamma p + B^{2})} \  F / \sqrt{\rho}$ is the local static slow continuum frequency.
These MHD continua form the basic organizing structure of the full MHD spectrum, and are schematically shown in Fig.~\ref{fig:spectrum}, 
where the equilibrium is assumed to be weakly inhomogeneous. The Eulerian entropy 
continuum (Goedbloed et al. \cite{GBHK})~$\Om{0} \equiv \vf{\kt}\cdot\vf{v}$ is also included in the figure.

\subsection{Local dispersion equation}
One can study many of the occuring instabilities by means of the local dispersion equation,
\begin{equation}
  \label{eq:dispersion}
    q^{2}\frac{A^{2}S^{2}}{r^{2}} + C^{2} + DE 
    - \frac{ASD}{r}\left( \frac{C}{D} \right)'
    = 0, 
\end{equation}
where $q$ is the local radial 'wavenumber'. This local dispersion equation can be derived from the differential 
\mbox{equations \eqref{eq:fod}} using the WKB method as discussed 
in appendix~\ref{sec:radialwavenumber}.
The equation will be solved neglecting the term proportional to $ASD/r$.  This approximation is valid if one 
assumes~$kc_{\mathrm{s}} \gg \omt$,  where~$c_{\mathrm{s}} \equiv \sqrt{\gamma p / \rho}$ is the sound speed.

The remaining local dispersion equation can be rewritten as a sixth-order polynomial in~$\omt$ which 
governs all discrete local modes. This dispersion equation reads: 
\begin{equation}
  \label{eq:polynomial}
  \omt^{6} + a_{4}\omt^{4} + a_{3}\omt^{3} + a_{2}\omt^{2} + a_{1}\omt + a_{0} = 0,
\end{equation}
where the coefficients~$a_{i}$ are defined as
\begin{align}
  \label{eq:a4}
  a_{4} & \equiv - \Biggl[ 
                (q^{2} + \kt^{2})\frac{\gamma p + B^{2}}{\rho} + \frac{F^{2}}{\rho} \\
        & \quad \qquad
                - \frac{r}{\rho} \left( \frac{B_{\theta}^{2}}{r^{2}} \right)' + \kappa^{2} + \frac{\rho'}{\rho}V_{g}
		\Biggr], \nonumber \\
  \label{eq:a3}
  a_{3} & \equiv - 4\Omega \left[ \frac{m}{r}V_{g} + 2\frac{kB_{z}B_{\theta}}{r\rho} \right], \\
  \label{eq:a2}
  a_{2} & \equiv (q^{2} + \kt^{2})\frac{2\gamma p + B^{2}}{\rho} \frac{F^{2}}{\rho} \\
        & \quad
                 + \kt^{2}\frac{\gamma p + B^{2}}{\rho} \Biggl[ 
		   - \frac{r}{\rho} \left( \frac{B_{\theta}^{2}}{r^{2}} \right)' + r{\Omega^{2}}' \nonumber \\
	& \quad \qquad \qquad \qquad \ 
		   + \frac{\rho'}{\rho}V_{g} - \frac{\rho V_{g}^{2}}{\gamma p + B^{2}} \Biggr] \nonumber \\
	& \quad
		 + 4k^{2}\Omega^{2}\frac{\gamma p + B^{2}}{\rho}
		 - 4k\left( \frac{m}{r}B_{z} - kB_{\theta} \right) \frac{B_{\theta}}{r\rho}V_{g} \nonumber \\
	& \quad
		 + 4\frac{\Omega^{2}F}{\rho} \left( \frac{m}{r}B_{\theta} - kB_{z} \right)
		 - 4\frac{k^{2}}{r^{2}} \frac{B_{\theta}^{2}B^{2}}{\rho^{2}},  \nonumber \\
  \label{eq:a1}		 
  a_{1} & \equiv 4 \frac{\Omega FB_{\theta}}{\rho} \left[ \kt^{2}V_{g} + 2k^{2}\frac{\gamma p}{r\rho} \right], \\
  \label{eq:a0}
  a_{0} & \equiv - \frac{\gamma p F^{2}}{\rho^{2}} \Biggl\{ 
                 (q^{2} + \kt^{2})\frac{F^{2}}{\rho} \\
        & \quad \qquad
                 + \kt^{2} \left[ 
		   - \frac{r}{\rho} \left( \frac{B_{\theta}^{2}}{r^{2}} \right)' + r{\Omega^{2}}'
		   + \frac{\rho'}{\rho}V_{g} - \frac{\rho V_{g}^{2}}{\gamma p} \right] \nonumber \\
	& \quad \qquad
		 - 4k^{2}\frac{B_{\theta}^{2}}{r^{2}\rho} \Biggr\}, \nonumber
\end{align}
with rotation frequency~$\Omega \equiv v_{\theta}/r$,  deviation from a Keplerian 
disk~$V_{g} \equiv v_{\theta}^{2}/r -  g$ and epicyclic frequency~$\kappa^{2} \equiv 2v_{\theta}(rv_{\theta})'/r^{2}$.
The local dispersion equation~\eqref{eq:dispersion} reduces to the local dispersion equation derived by
Dubrulle and Knobloch (\cite{DK}) if one considers axisymmetric perturbations, a constant density, a zero radial
`wavenumber' $q$ and takes the incompressible limit.

\subsection{Magneto-rotational instability}
By making some extra assumptions, we can reduce the local dispersion equation~\eqref{eq:dispersion} further.
Consider {\it axisymmetric} perturbations and the situation where $qc_{s}$, $kc_{s}$, $qv_{\mathrm{A}}$ 
and $kv_{\mathrm{A}} \gg \omt$ and a purely axial magnetic field. Here, $v_{\mathrm{A}} \equiv B_{z}/\sqrt{\rho}$ is 
the Alfv\'en speed. In that case, the local dispersion equation~\eqref{eq:polynomial} reduces to the following
$4^{\rm th}$ order polynomial: 
\begin{equation}
  \label{eq:polynomialBH}
  b_{4}\omt^{4} + b_{2}\omt^{2} + b_{0} = 0,
\end{equation}
where the coefficients are 
\begin{align}
  b_{4} & = (q^{2} + k^{2}) \frac{\gamma p + B_{z}^{2}}{\rho} + \frac{F^{2}}{\rho}, \\
  b_{2} & = - \Biggl\{ (q^{2} + k^{2})\frac{2\gamma p + B_{z}^{2}}{\rho} \frac{F^{2}}{\rho} \\
        & \quad \qquad \
                      + k^{2}\frac{\gamma p+B_{z}^{2}}{\rho} 
		      \left[ r{\Omega^{2}}' + \frac{\rho'}{\rho}V_{g} - \frac{\rho V_{g}^{2}}{\gamma p + B_{z}^{2}} 
		      \right] \nonumber \\
	& \quad \qquad \
		      + 4k^{2}\Omega^{2} \frac{\gamma p}{\rho} \Biggr\}, \nonumber \\
  b_{0} & = \frac{\gamma p F^{2}}{\rho^{2}} \Biggl\{
            (q^{2} + k^{2})\frac{F^{2}}{\rho} \\
        & \quad \qquad \qquad
            + k^{2}\left[ r{\Omega^{2}}' + \frac{\rho'}{\rho}V_{g} - \frac{\rho V_{g}^{2}}{\gamma p} \right] \Biggr\}.
	    \nonumber
\end{align}
This $4^{\rm th}$ order polynomial can be solved analytically, which results in the stability criterion
\begin{equation}
  \label{eq:stabilityBH}
  \frac{F^{2}}{\rho} \ge \frac{-k^{2}}{q^{2}+k^{2}} \left[ 
                         r {\Omega^{2}}' + \frac{\rho'}{\rho}V_{g} - \frac{\rho V_{g}^{2}}{\gamma p} \right].
\end{equation}
This is the stability criterion for the axisymmetric MRI for disks with a purely axial magnetic field. 
This criterion is more generally applicable than the one derived by Balbus and Hawley (\cite{BH})
to which it reduces for a weakly magnetized disk ($p \gg B_{z}^{2}$).

\section{The accretion disk model in the cylindrical limit}
In order to quantify instabilities using a linear analysis, one has to
consider the equilibrium state of an accretion disk. The model we consider 
uses power-law scalings for the different flow variables, following the 
self-similar models of Spruit et al. (\cite{SMIS}). In order to have a model 
that is in an equilibrium state, these profiles have to satisfy equation (11).
We use the following profiles for the density $\rho$, 
thermal pressure $p$, toroidal magnetic field $B_{\theta}$, axial magnetic 
field $B_{z}$ and the toroidal velocity $v_{\theta}$:
\begin{align}
   \rho       & = r^{-3/2},                 \\
   p          & = \epsilon^2 \;\; r^{-5/2}, \\
   B_{\theta} & = -\alpha_1
                  \sqrt{{2\epsilon^2\over\beta(\alpha_1^2+\alpha_2^2)}}\;\; r^{-5/4}, \\
   B_{z}      & = \alpha_2
                  \sqrt{{2\epsilon^2\over\beta(\alpha_1^2+\alpha_2^2)}}\;\; r^{-5/4}, \\
   v_{\theta} & =  V_0\;\; r^{-1/2},        \\
\intertext{where the $\alpha$-parameters express the ratio of toroidal or
axial to the total magnetic field,}
   \frac{B_{\theta}}{B} & = \frac{-\alpha_{1}}{\sqrt{\alpha_{1}^{2} + \alpha_{2}^{2}}}, \\
   \frac{B_{z}}{B}      & = \frac{ \alpha_{2}}{\sqrt{\alpha_{1}^{2} + \alpha_{2}^{2}}},
\end{align}
and
\begin{equation}
   V_0^2 =  GM_{*} - {\epsilon^2\over 2\beta(\alpha_1^2 + \alpha_2^2)}
            \bigl[ 5(1+\beta)(\alpha_1^2 + \alpha_2^2) - 4\alpha_1^2 \bigr].
\end{equation}
Finally, the parameter $\beta$ measures the radially constant ratio between
the thermal pressure and the total magnetic pressure , i.e. 
$\beta\equiv 2p/B^2$. Notice that the density and pressure 
profiles are such that the entropy is constant throughout the disk, thereby
excluding convective modes.

One of the key parameters of our model is the parameter $\epsilon$,
which is taken as a constant free parameter. For values of $\epsilon \ll 1$, 
the physical interpretation is the same as in the model of Shakura \& Sunyaev 
(\cite{SS}), i.e. $\epsilon=c_{\rm s}/v_{\theta}\simeq H/r$, where~$H$
is the scale height of the disk. These cases correspond to thin Keplerian 
rotating accretion disks. For values of $\epsilon\sim 1$, one is in the 
regime of sub-Keplerian rotating disks (see e.g. Narayan \& Yi \cite{NY}) 
and for these cases we stick to the interpretation $\epsilon\simeq H/r$, 
noting that the identification $\epsilon=c_{\rm s}/v_{\theta}$ is not valid anymore. 
Typically $c_{\rm s}/v_{\theta}$ will be larger, up to a maximum factor of order $\sim 10$.

Models of accretion disks which only depend on the radius $r$ are sometimes
referred to as accretion disks in the cylindrical limit (see for example 
Armitage \cite{A} and Hawley \cite{H}). The model we use is an example of such a model, 
where the results from these models approximate the interior equatorial 
region of the accretion disk. This approximation is correct as long as the axial wavenumber~$k$ 
is much larger than the inverse of the scale height $H$. Therefore the 
identity $k \gg 2\pi /(\epsilon r)$ has to be satisfied in order to connect the results 
from our spectral analysis with MHD stability properties of the interior of an accretion disk.

\section{Numerical codes}
For the investigation of the stability properties two numerical codes, LEDAFLOW and a
Local Dispersion Equation Solver (LODES), have been used. In the next two subsections 
we briefly explain their algorithmic details.

\subsection{The spectral code LEDAFLOW}
The code LEDAFLOW developed by Nijboer et al. (\cite{NHPG}) can compute
the complete MHD spectrum for a given 
one-dimensional stationary equilibrium, axial wavenumber $k$, and azimuthal 
wavenumber $m$. Instead of numerically solving the system~\eqref{eq:fod} the full
set of  MHD equations \eqref{eq:momentum}-\eqref{eq:density} 
are linearized and then discretized using finite elements in the 
inhomogeneous direction, and solved using the Galerkin method 
(Schwarz \cite{S}). For the elements, a combination of cubic Hermite 
and quadratic elements for the perturbed quantities is used to prevent the 
creation of spurious eigenvalues. The result is a non-Hermitian eigenvalue problem, 
which is solved using a QR method or an inverse vector 
iteration method (Press et al. \cite{PTVF}). The latter one can be used to calculate the 
eigenfunctions for a single eigenvalue. The used boundary conditions treat 
the edges of the disk as perfect conducting walls. Strictly speaking,
these boundary conditions are inappropriate for considering waves modes in an accretion disk. 
However, interior local modes that do not depend strongly on the precise variations
in a boundary region are barely affected by these conditions.

\subsection{The local dispersion equation solver}
The local dispersion equation solver LODES directly computes the roots 
of the local dispersion equation~\eqref{eq:polynomial} at a given radial position. 
This is done by  making use of Laguerre's method (Press et al. \cite{PTVF}). 

The code calculates the local value of coefficients $a_{i}$~\eqref{eq:a4}-\eqref{eq:a0}
depending on the given equilibrium, the radial wavenumber $q$, the azimuthal 
wavenumber $m$ and the axial wavenumber $k$. In principle the code 
can compute all six roots, but there is a switch to calculate only the
mode with the largest growth rate. Because the code uses Laguerre's method 
(Press et al. \cite{PTVF}), it needs an initial guess for this root. As an
initial guess, we use the analytical solution of the most unstable mode of 
the $4^{\mathrm{th}}$ order polynomial~\eqref{eq:polynomialBH}.

\subsection{Coupling LEDAFLOW with LODES \label{subsec:coupling}}
As described in the previous subsection, LODES needs a value of the radial 
'wavenumber' $q$ and a radial position $r_{\rm i}$ to calculate
the roots. In this subsection, we show how we extract these input parameters
from the results of LEDAFLOW. 

\begin{figure}
  \centering
  \epsfig{file=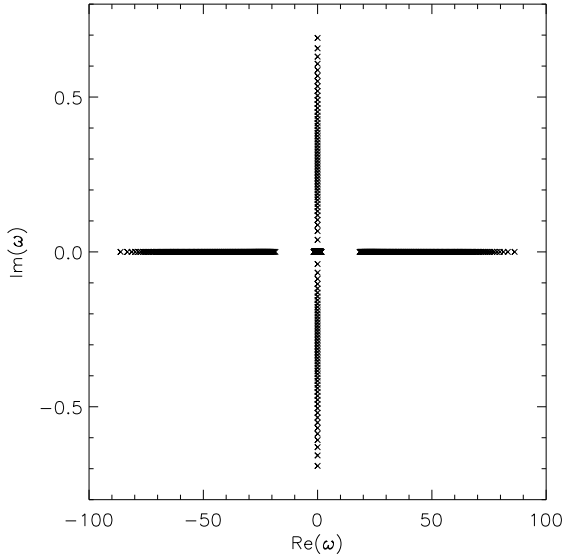, width=8cm}
  \caption{Complete MHD spectrum of a weakly magnetized thin disk for wavenumbers~$m=0$ and $k=200$.}
  \label{fig:mrispectrum}
\end{figure}

Fig.~\ref{fig:mrispectrum} shows the complete MHD spectrum of a
calculation performed by LEDAFLOW with the following set of parameters:
$m=0$, $k=200$, $\beta = 1000$, $\epsilon = 0.1$ and a purely 
axial magnetic field ($\alpha_{1} = 0$ and $\alpha_{2} = 1$).
In this calculation, we use 101 grid points in the radial direction on the
domain $r=[1,2]$.
It is clear from the resulting MHD spectrum that the disk is not stable.
We identify these {\it unstable} modes as MRI modes. 
Notice that, for a purely axial magnetic field, the eigenvalues~$\omega^{2}$ are real
(i.e. no overstability in this case).

\begin{figure}
  \centering
  \epsfig{file=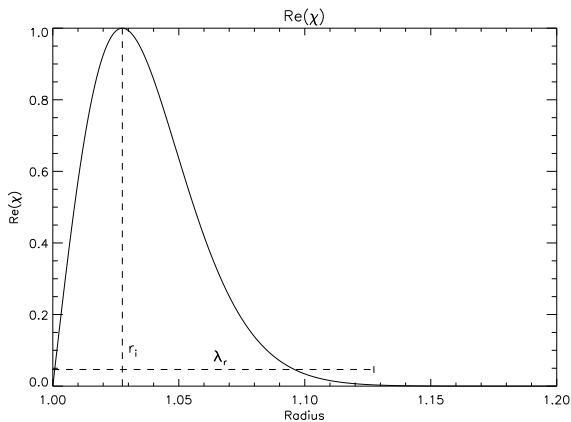, width=8cm}
  \caption{The eigenfunction $\chi$ of the mode with the largest growth rate.
           The growth rate is $0.728\Omega$ .}
  \label{fig:mrichi}
\end{figure}

The eigenfunction $\chi$ of the mode with the largest growth rate is shown 
in Fig.~\ref{fig:mrichi}. 
From this eigenfunction the location $r_{\mathrm{i}}$ and the radial 
wavenumber~$q$ of the instability can be extracted. The location 
$r_{\mathrm{i}}$ is the radius where $\chi$ has its extremum. The radial 
wavenumber~$q$ follows from~$q = 2\pi / \lambda_{r}$, 
where $\lambda_{r}$ is chosen as the radial wavelength from the inner boundary 
to the point when the value of the eigenfunction is 0.3\% of its extremum value. 
The justification of this heuristic method to calculate the radial `wavenumber'~$q$ 
can be found in the appendix, where the WKB solutions are matched to the analytical 
solutions in the turning point regions.

The mode shown in Fig.~\ref{fig:mrichi} has also been investigated on the enlarged
radial domains $r=[0.966,2]$, $r=[0.938,2]$ and $r=[0.915,2]$. The inner boundary of 
these domains is such that the mode has its extremum at the same radial location for the 
different radial domains. The resulting growth rates are all almost equal to each other, 
which confirms our earlier statement that the mode does not strongly depend on the 
boundary conditions used.

\section{The magneto-rotational instability}
We present results on magneto-rotational instabilities in
an accretion disk in the cylindrical limit using the methods described in the 
previous section. The calculations using LEDAFLOW are performed on a radial 
domain $r=[1,2]$. We only consider axisymmetric perturbations ($m = 0$) and 
take a ratio of specific heats~$\gamma = 5/3$. We discuss seven calculations
of which the different sets of parameters are shown in Tables~\ref{tab:Bt} and~\ref{tab:Vg}.  

\subsection{Influence of the toroidal magnetic field}
First, we investigate the influence of the toroidal magnetic field on both the 
growth rate and the oscillation frequency of the MRI. This is done by varying the
toroidal magnetic field strength, while keeping the axial magnetic field strength 
constant. This means that the plasma beta varies between the different calculations.
The parameters of the four calculations are shown in Table~\ref{tab:Bt}.
In these calculations we consider a thin disk by taking $\epsilon = 0.1$. 
\begin{table}
  \centering
  \caption{Parameters of thin, Keplerian accretion disks}
  \begin{tabular}{|c|c|c|c|c|c|c|}
    \hline
    Sim. no. & $\epsilon$ & $\alpha_{1}$ & $\alpha_{2}$ & $\beta$ & $k$     & $\Delta k$  \\ \hline
    A        & 0.1        & 0            & 1            & 1000    & 200-365 & 5           \\ \hline
    B        & 0.1        & 1            & 1            & 500     & 200-365 & 5           \\ \hline
    C        & 0.1        & 10           & 1            & 9.901   & 200-365 & 5           \\ \hline
    D        & 0.1        & 30           & 1            & 1.110   & 200-365 & 5           \\ \hline
  \end{tabular}
  \label{tab:Bt}
\end{table}

\begin{figure*}
  \centering
  \begin{tabular}{cc}
    \epsfig{file=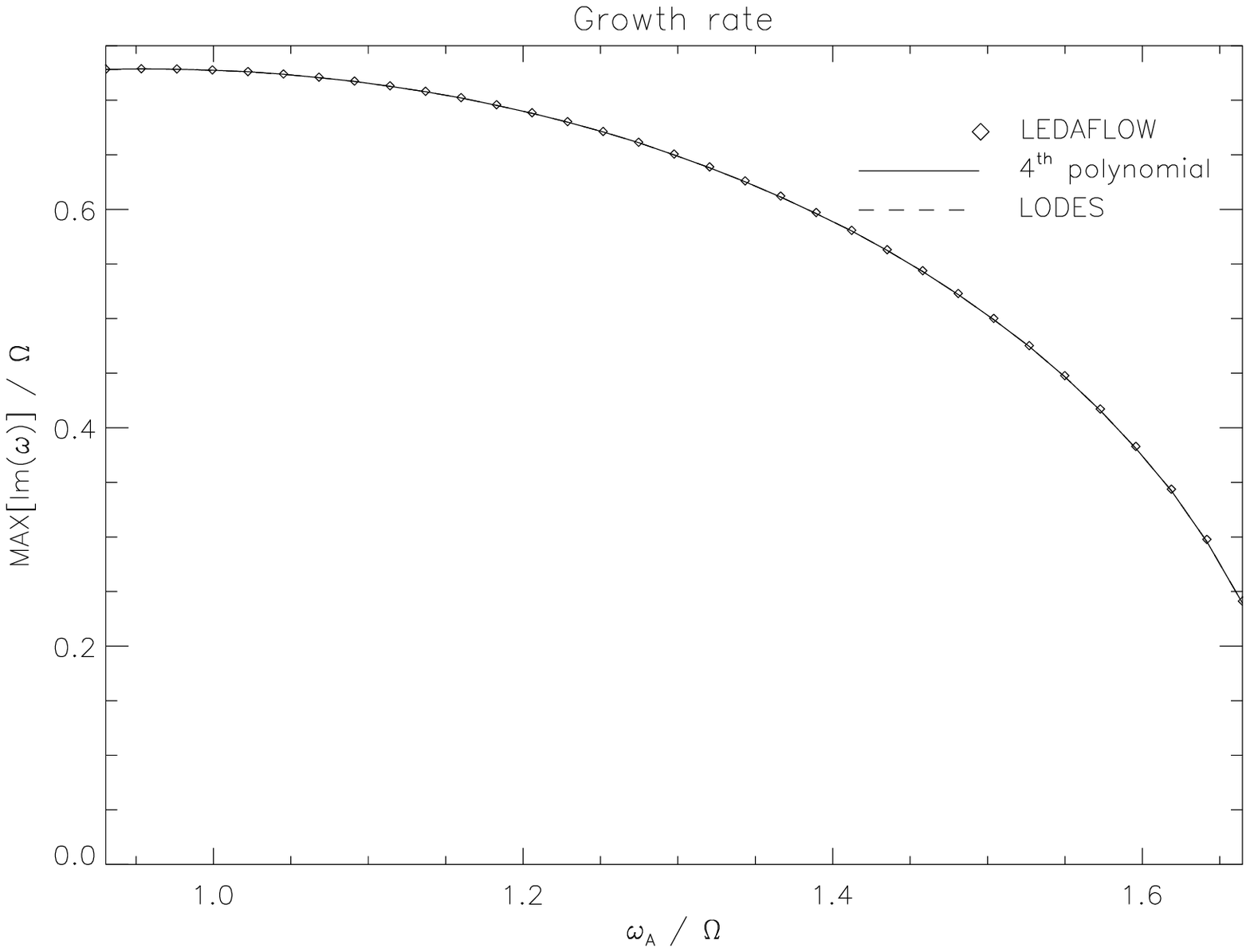, width=8cm} &
    \epsfig{file=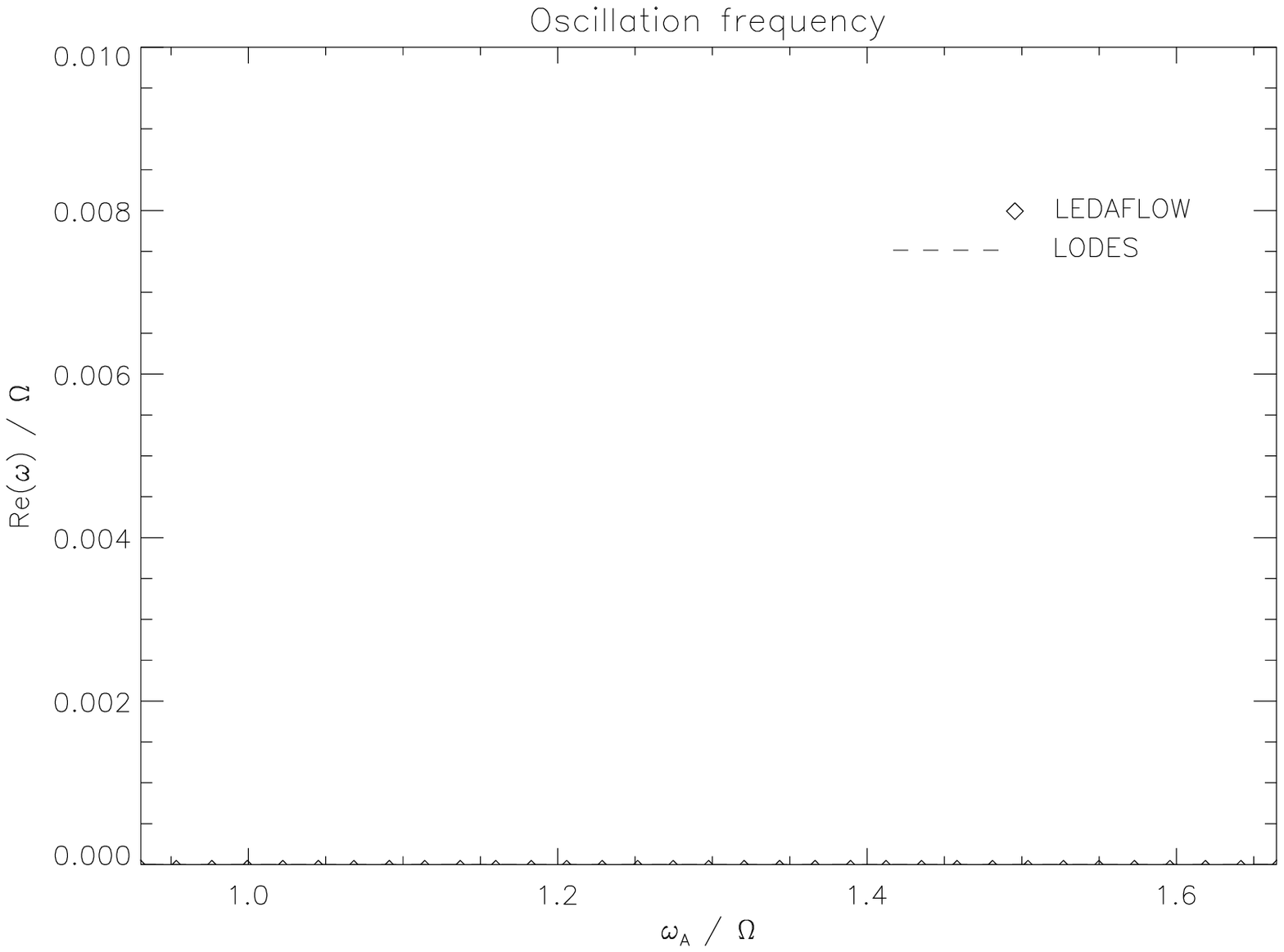, width=8cm}
  \end{tabular}
  \caption{The growth rate and the oscillation frequency of the most unstable axisymmetric MRI mode
           as a function of the scaled Alfv\'en frequency for disk model A, where $\beta = 1000$.
           In this purely axial magnetic field configuration, the axisymmetric modes are purely
           exponentially growing in time.}
  \label{fig:Bt1}
\end{figure*}

\begin{figure*}
  \centering
  \begin{tabular}{cc}
    \epsfig{file=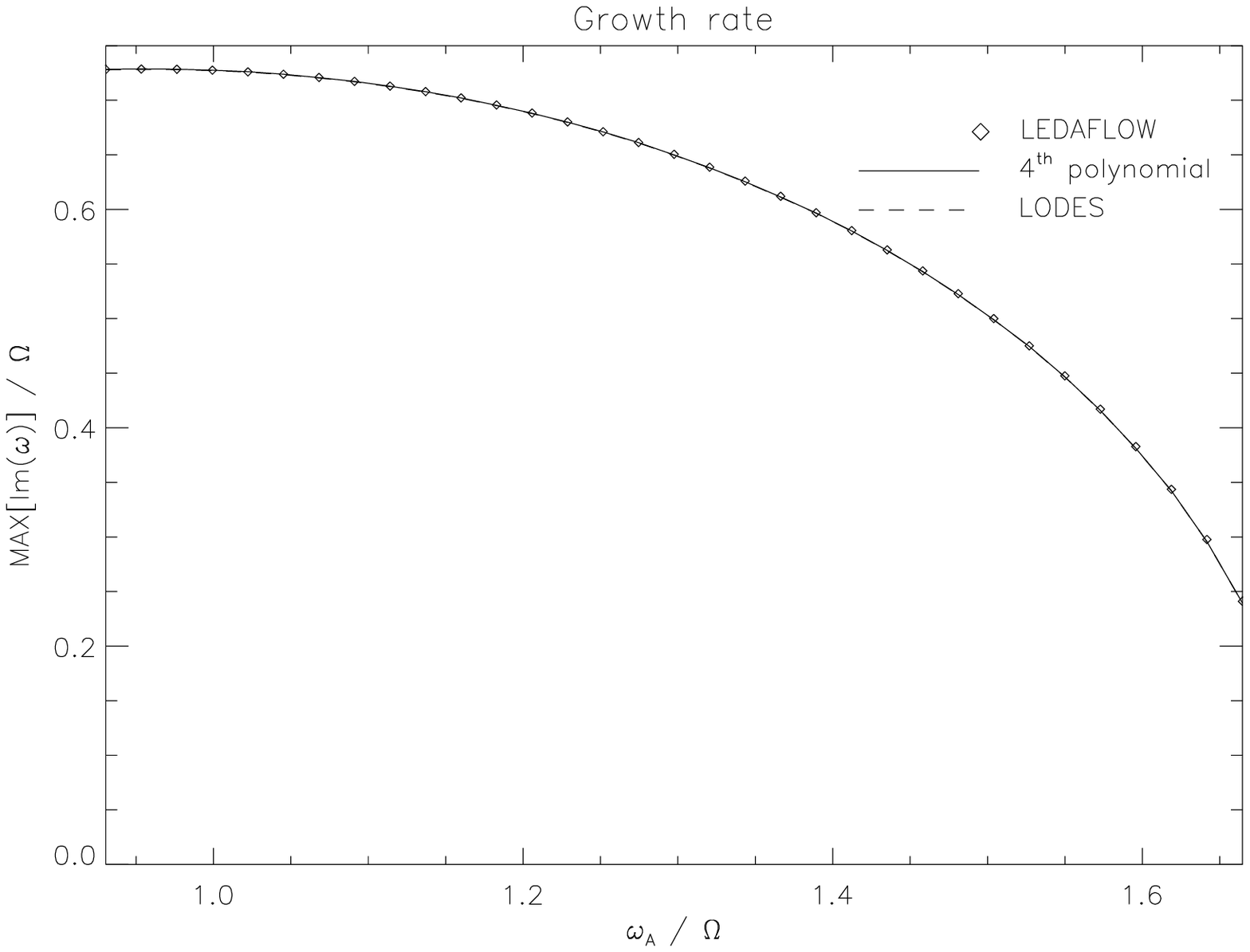, width=8cm} &
    \epsfig{file=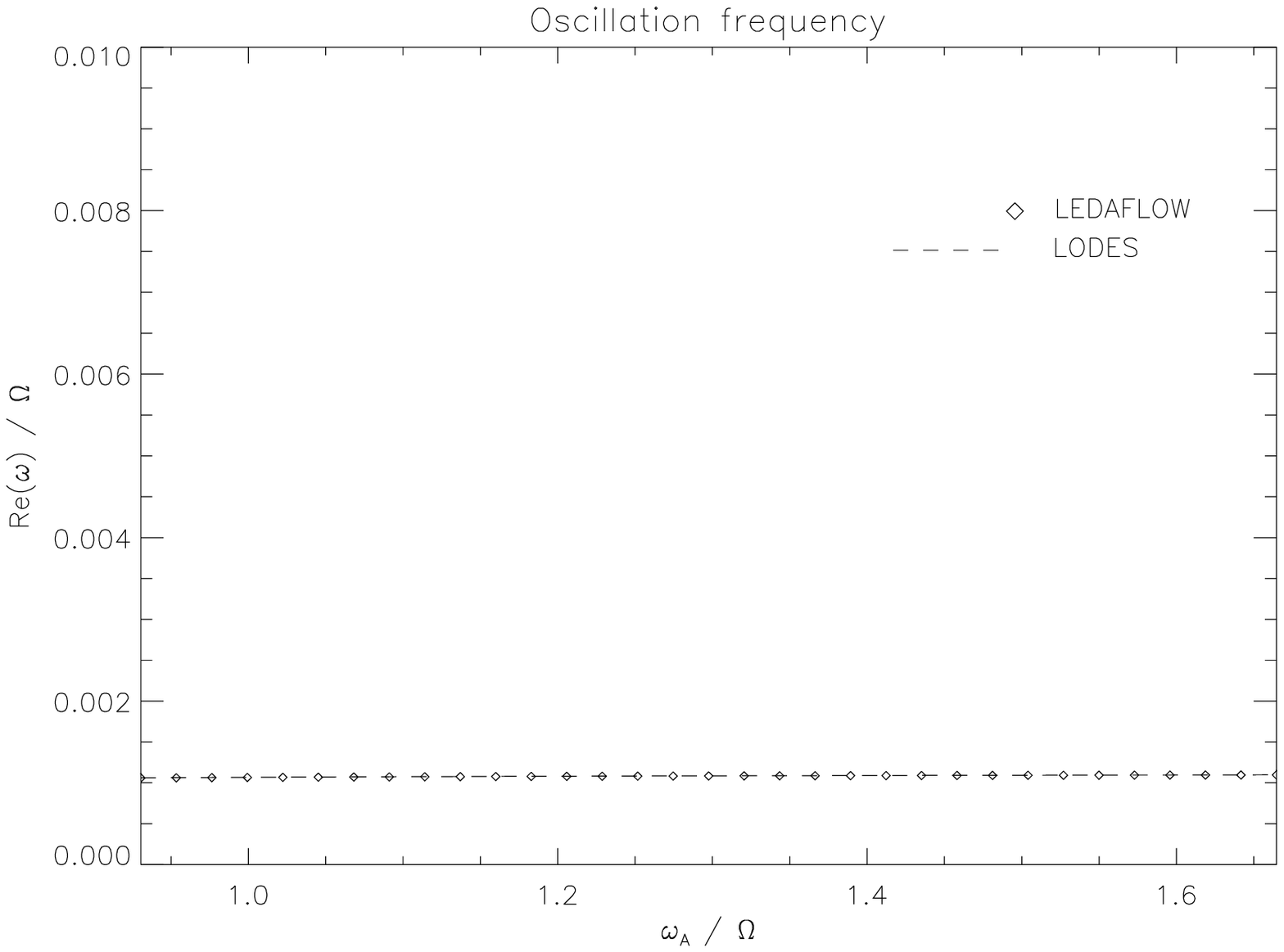, width=8cm}
  \end{tabular}
  \caption{The growth rate and the oscillation frequency of the most unstable axisymmetric MRI mode
           as a function of the scaled Alfv\'en frequency for disk model B, where $\beta = 500$.}
  \label{fig:Bt2}
\end{figure*}

\begin{figure*}
  \centering
  \begin{tabular}{cc}
    \epsfig{file=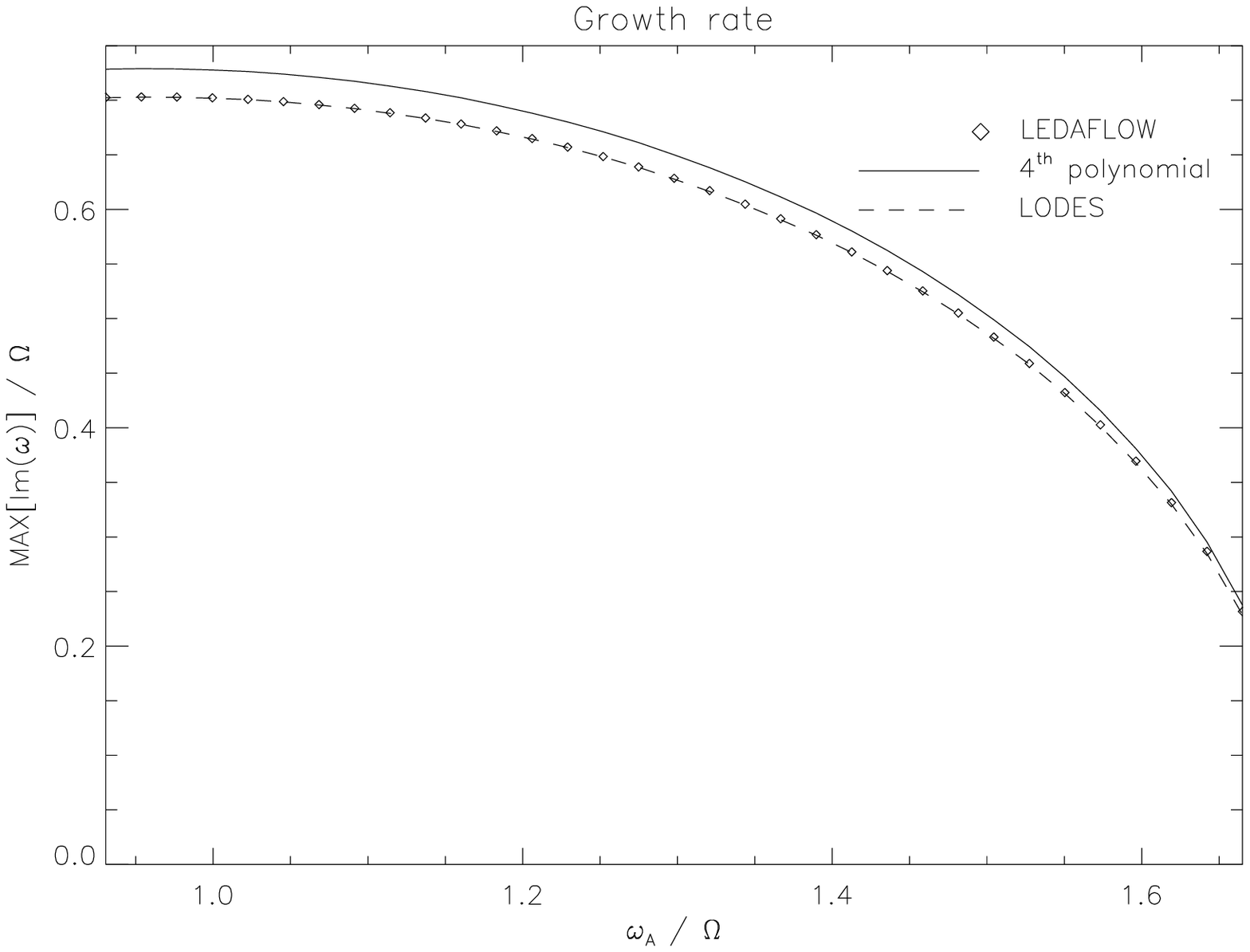, width=8cm} &
    \epsfig{file=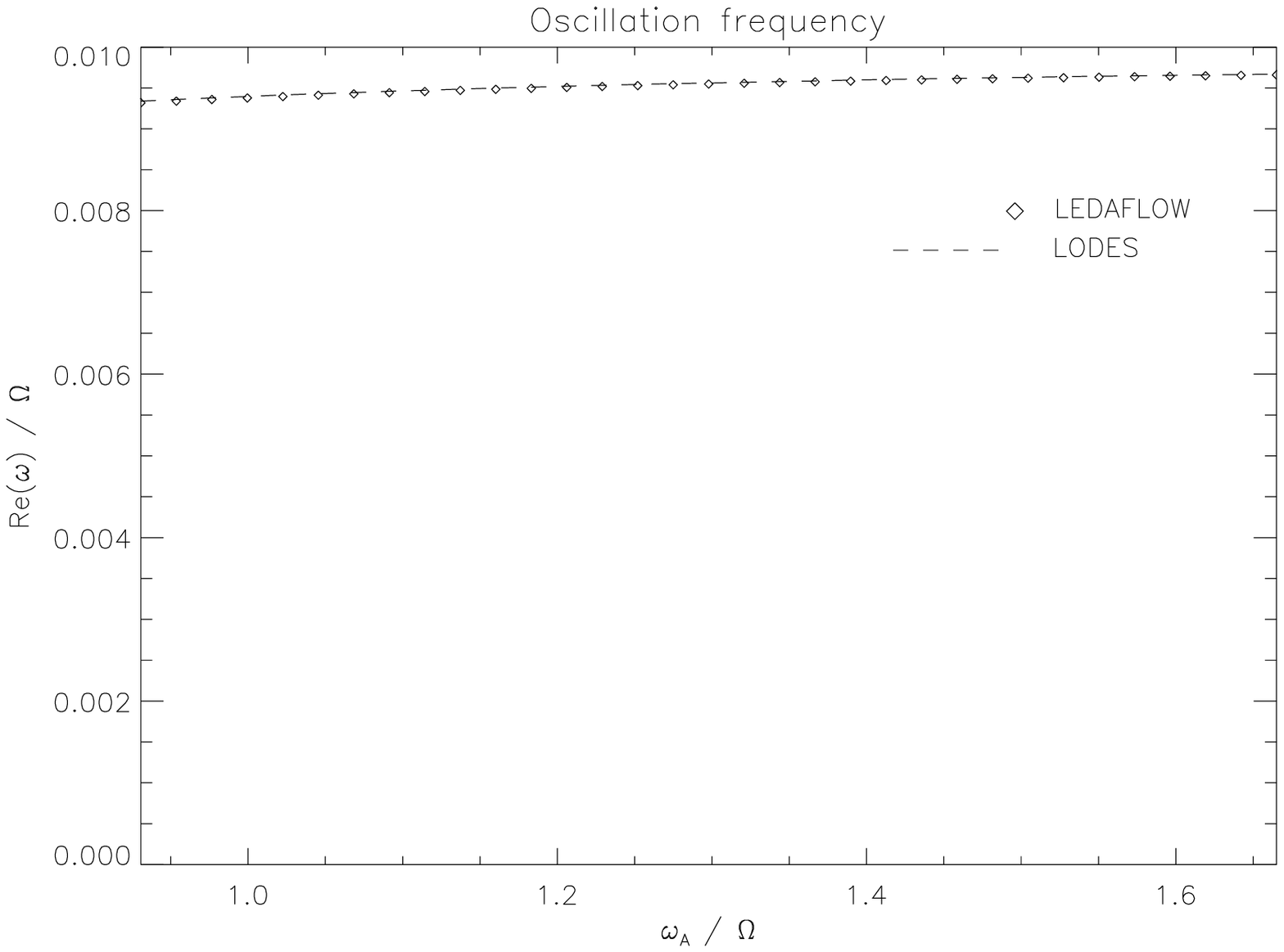, width=8cm}
  \end{tabular}
  \caption{The growth rate and the oscillation frequency of the most unstable axisymmetric MRI mode
           as a function of the scaled Alfv\'en frequency for disk model C, where $\beta = 9.901$.}
  \label{fig:Bt3}
\end{figure*}

\begin{figure*}
  \centering
  \begin{tabular}{cc}
    \epsfig{file=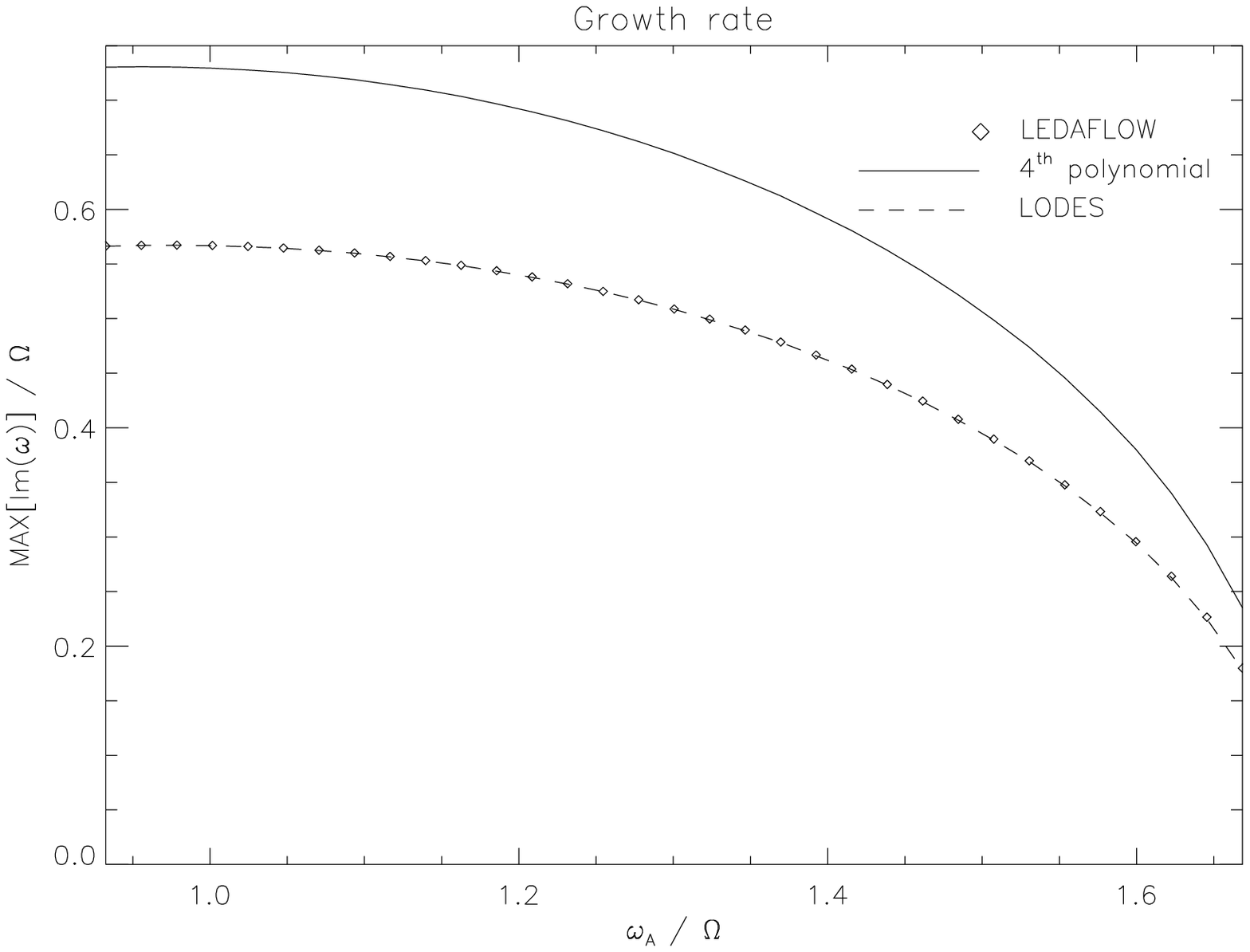, width=8cm} &
    \epsfig{file=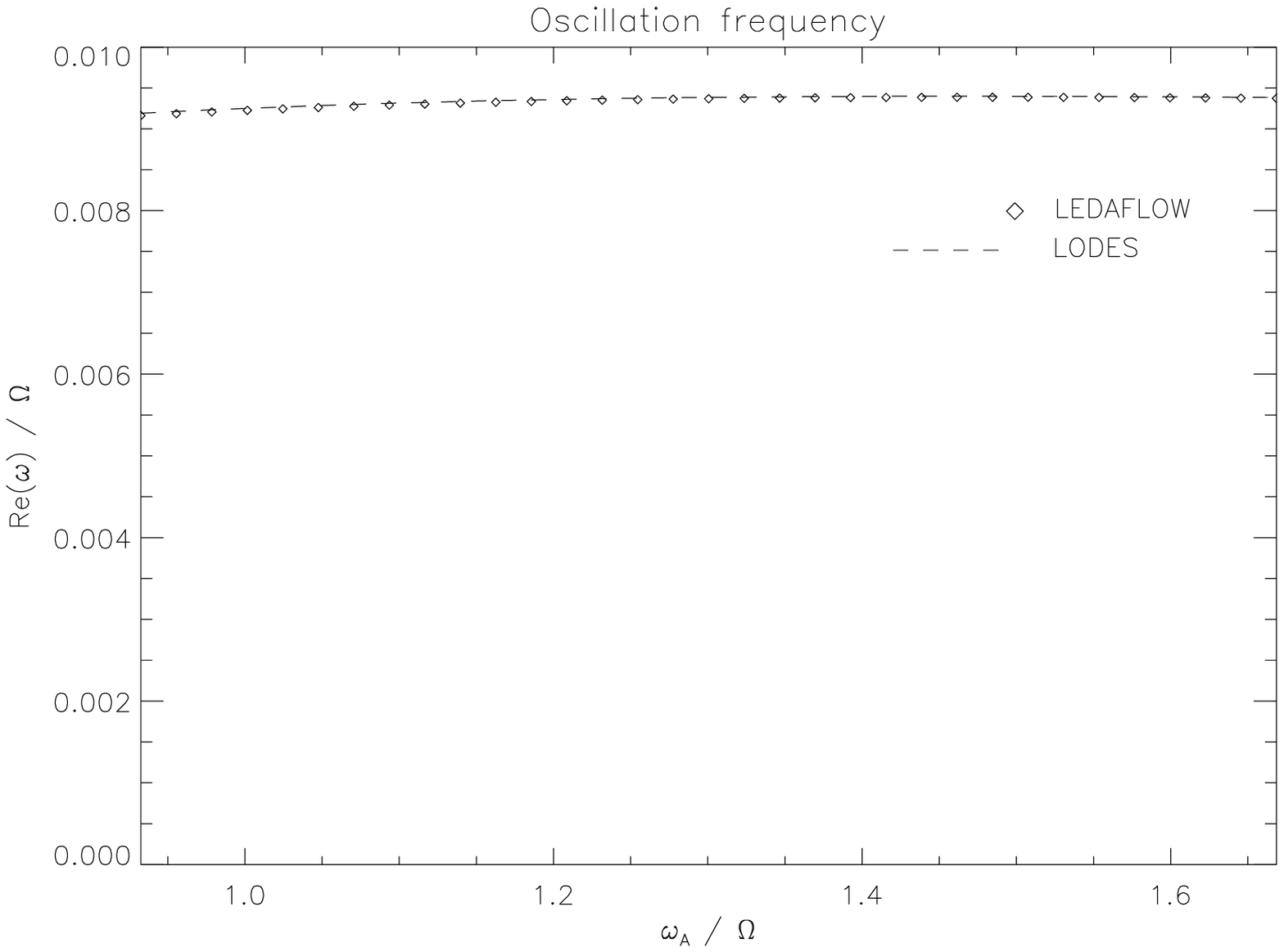, width=8cm}
  \end{tabular}
  \caption{The growth rate and the oscillation frequency of the most unstable axisymmetric MRI mode
           for a disk close to equipartition with a dominant toroidal magnetic field component
	   (disk model D).}
  \label{fig:Bt4}
\end{figure*}

The results are shown in Figs.~\ref{fig:Bt1}-\ref{fig:Bt4}. In
each figure, the left panel shows the growth rate and the right panel shows the oscillation
frequency of the most unstable axisymmetric mode, both as a function of the axial wavenumber. 
We have scaled the growth rate and the oscillation frequency with respect to the rotation frequency. 
Instead of the axial wavenumber $k$, we use the corresponding local Alfv\'en frequency~$\omega_{\mathrm{A}}$
scaled with respect to the local rotation frequency. This scaling is similar to the one used by 
Narayan et al. (\cite{NQIA}).

In our figures, the diamonds represent the calculations performed with LEDAFLOW. The dashed
and solid line correspond with the results from respectively the analytical solution 
of the approximate $4^{\rm th}$ order polynomial~\eqref{eq:polynomialBH} and LODES. 
Recall that the last two calculations make use of LEDAFLOW results, in the manner described in 
section \ref{subsec:coupling}. 

In the calculations A, B, and C (shown in Figs.~\ref{fig:Bt1}, \ref{fig:Bt2}, and~\ref{fig:Bt3}, 
respectively) we change from a purely axial (A) to a predominantly toroidal (C) magnetic field configuration.
One sees that the growth rate of the analytical solution of 
$4^{\rm th}$ order polynomial~\eqref{eq:polynomialBH} and LODES matches with the LEDAFLOW calculations. 
The differences to LEDAFLOW are less than 4\% and 1\%, respectively.  These calculations (A, B and C) 
also show that the oscillation frequency increases away from zero as the 
toroidal magnetic field strength increases. A similar result was obtained by Dubrulle and 
Knobloch (\cite{DK}). These authors considered an {\it incompressible} plasma with a similar toroidal velocity
profile as used in this paper, but a different toroidal magnetic field profile.
Again, there is a perfect match between the LEDAFLOW and LODES results. However, the $4^{\rm th}$ order 
polynomial~\eqref{eq:polynomialBH} cannot be used to compute the oscillation frequency since it only  yields a value 
for the growth rate. The differences between the LEDAFLOW and LODES results are again less than 1\%.  

Fig.~\ref{fig:Bt4} shows results from a calculation for which the accretion disk
is close to equipartition in a strong toroidal~$|B_{\theta}|=30|B_{z}|$ magnetic field configuration
(disk model D). Again, there is a perfect match between the results from LODES and 
LEDAFLOW (the discrepancy is less than 1\%), but the $4^{\mathrm{th}}$ order polynomial~\eqref{eq:polynomialBH}
no longer predicts a correct approximation of the growth rate. This is due to the dynamical importance of the
toroidal magnetic field component. Comparing Fig.~\ref{fig:Bt4} with Fig.~\ref{fig:Bt3}, it is clear that
the oscillation frequency remains roughly the same and the growth rate decreases slightly. 
The analytical work performed by Dubrulle and Knobloch (\cite{DK}) also indicates a decrease in
the growth rate, as a result of the increasing tension of the toroidal magnetic field.
Hence, the cases A-D clearly demonstrate that the MRI remains active when the toroidal magnetic field increases. 
This remains true at least up to equipartition (Fig.~\ref{fig:Bt4}).
 
\subsection{Keplerian to sub-Keplerian disk}
In this subsection, we investigate the properties of the axisymmetric MRI in sub-Keplerian 
disks. Three calculations have been performed, all with $\beta=1000$, 
$\alpha_{1}=10$ and $\alpha_{2}=1$ (see Table~\ref{tab:Vg}). A sub-Keplerian 
disk can be obtained by increasing the value of the $\epsilon$ parameter beyond
$\epsilon =0.3$. All calculations have been performed such that the domain of the 
Alfv\'en frequency over the rotation frequency remains the same: the corresponding 
domain of $k$-values can be found in Table~\ref{tab:Vg}.
\begin{table}
  \centering
  \caption{Parameters of Keplerian to sub-Keplerian accretion disks}
   \begin{tabular}{|c|c|c|c|c|c|c|}
    \hline
    Sim. no. & $\epsilon$ & $\alpha_{1}$ & $\alpha_{2}$ & $\beta$ & $k$  & $\Delta k$ \\ \hline
    E        & 0.1        & 10           & 1            & 1000    & 320-4000 & 40             \\ \hline
    F        & 0.3        & 10           & 1            & 1000    & 96-1200  & 12             \\ \hline
    G        & 0.5        & 10           & 1            & 1000    & 40-500   & 5              \\ \hline
  \end{tabular}
  \label{tab:Vg}
\end{table}

\begin{figure*}
  \centering
  \begin{tabular}{cc}
    \epsfig{file=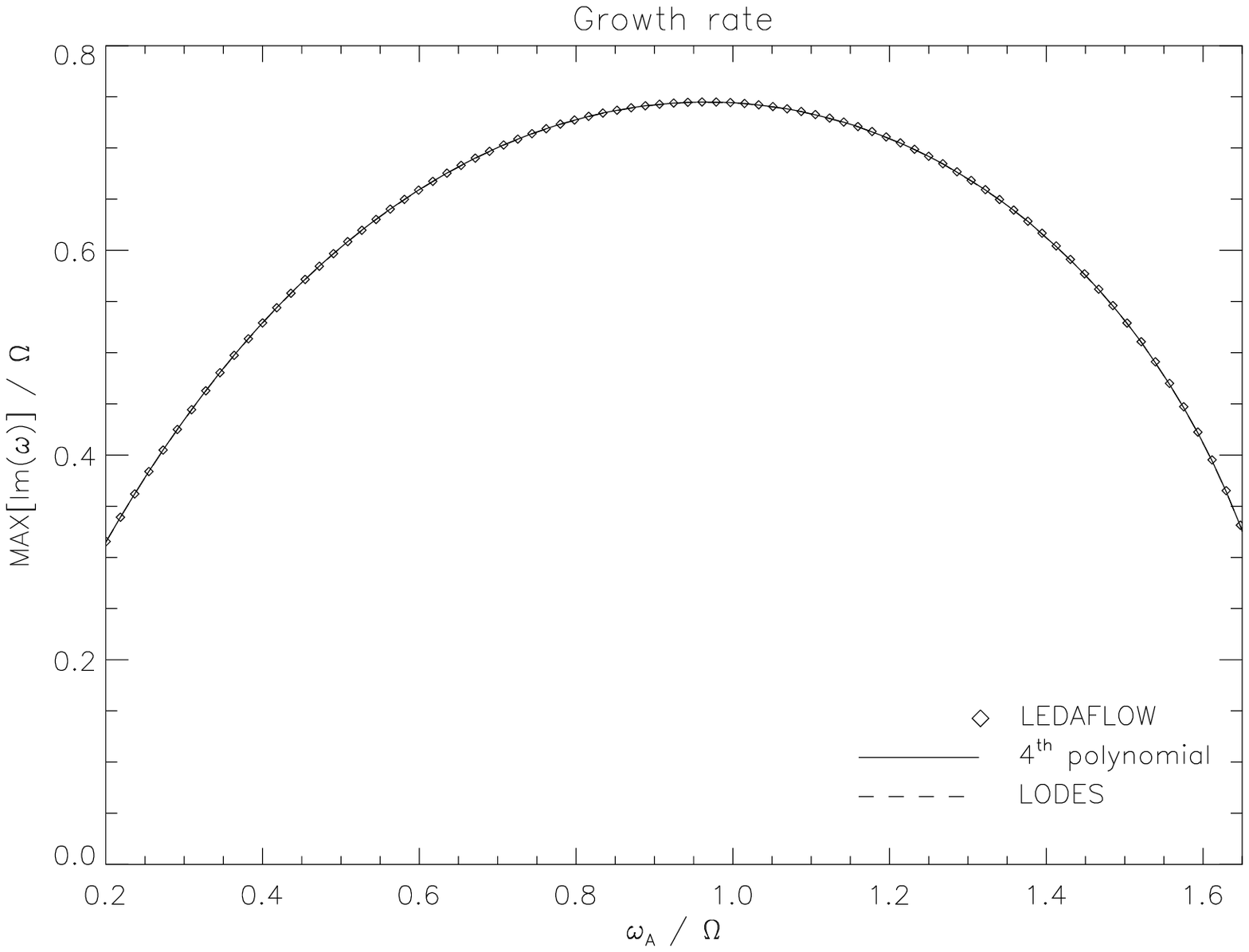, width=8cm} &
    \epsfig{file=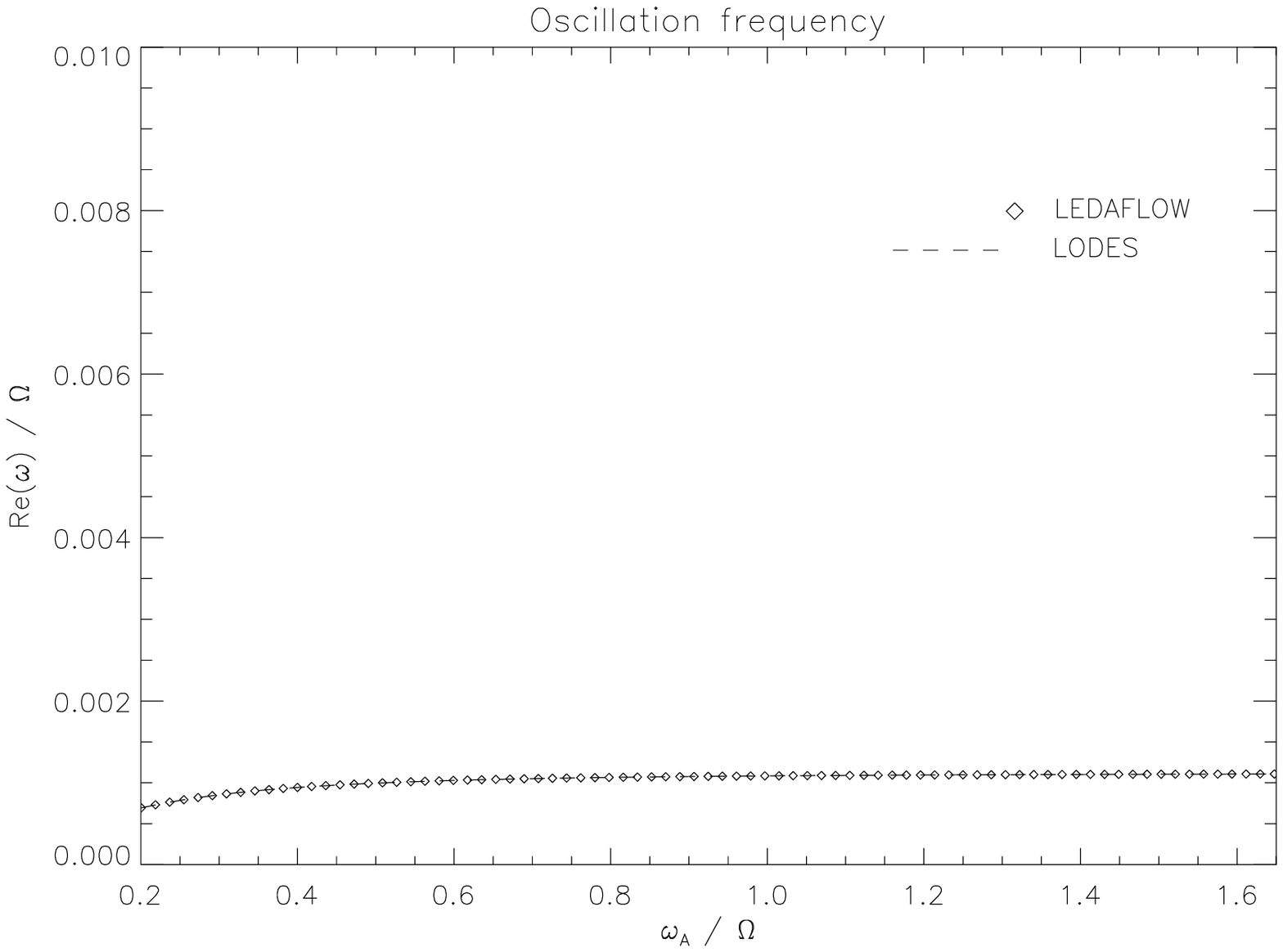, width=8cm}
  \end{tabular}
  \caption{The growth rate and the oscillation frequency of the most unstable mode 
           for disk model E, where the disk rotation is Keplerian.}
  \label{fig:Vg1}
\end{figure*}

\begin{figure*}
  \centering
  \begin{tabular}{cc}
    \epsfig{file=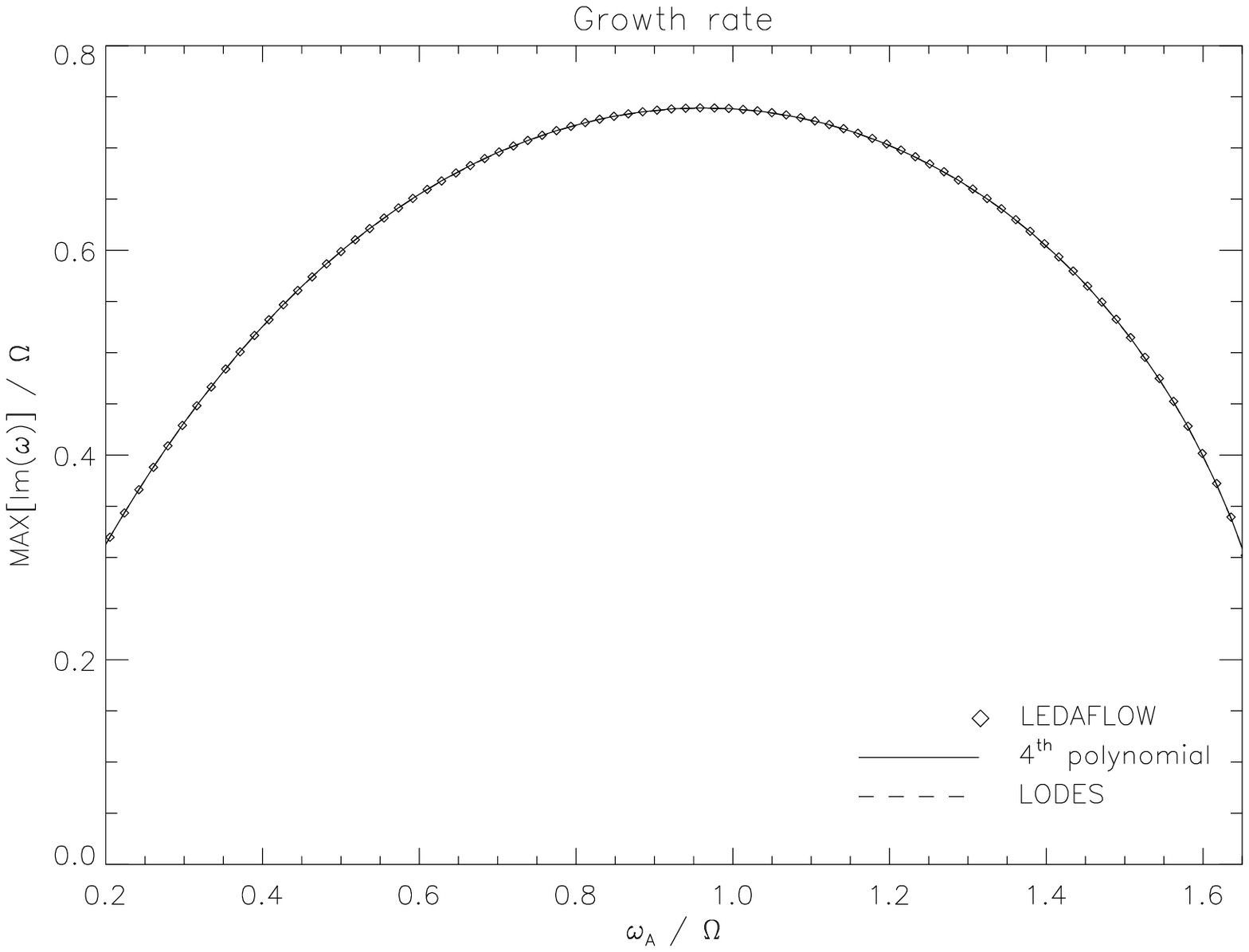, width=8cm} &
    \epsfig{file=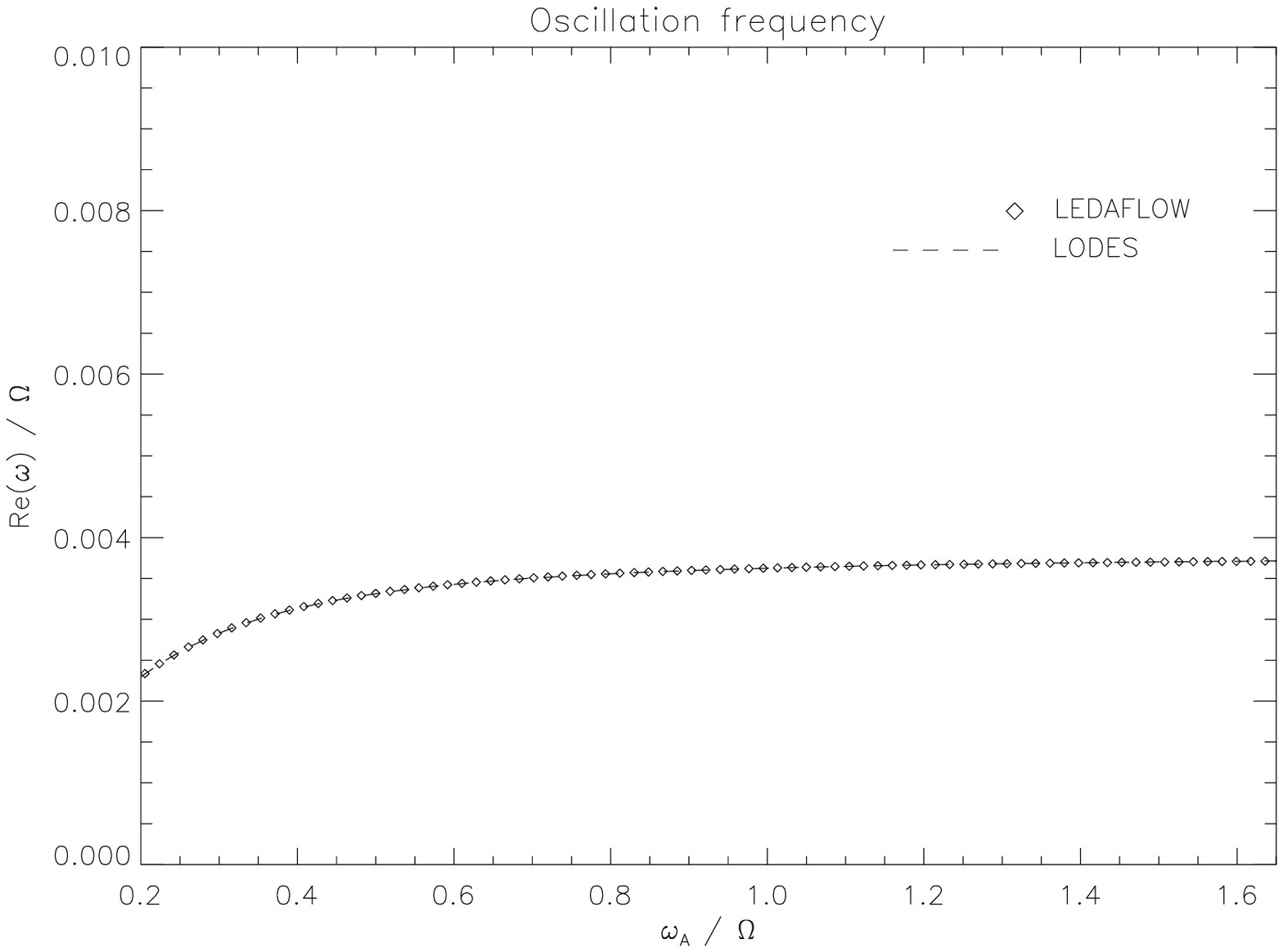, width=8cm}
  \end{tabular}
  \caption{The growth rate and the oscillation frequency of the most unstable mode 
           for disk model F, where the disk rotation is weakly sub-Keplerian.}
  \label{fig:Vg2}
\end{figure*}

\begin{figure*}
  \centering
  \begin{tabular}{cc}
    \epsfig{file=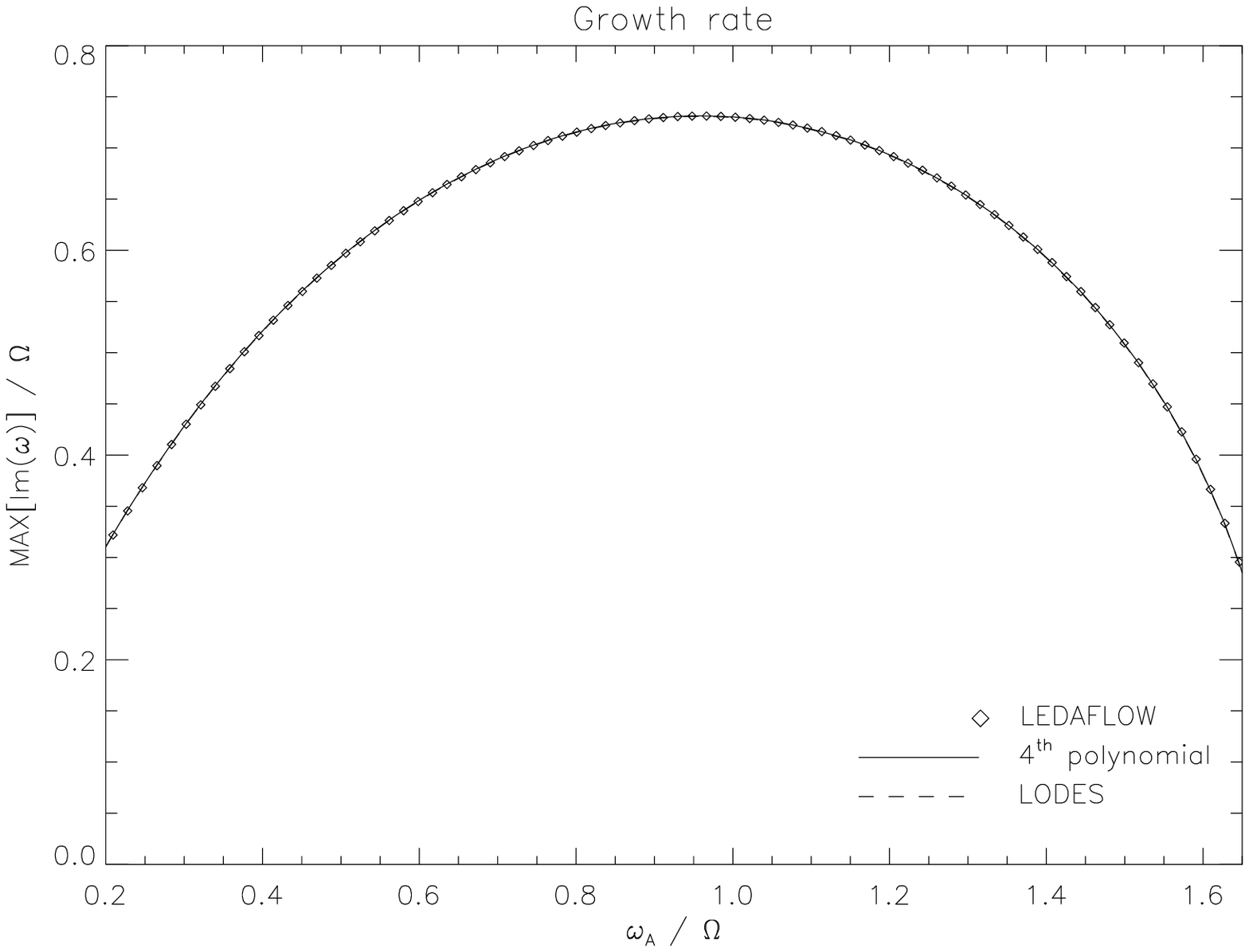, width=8cm} &
    \epsfig{file=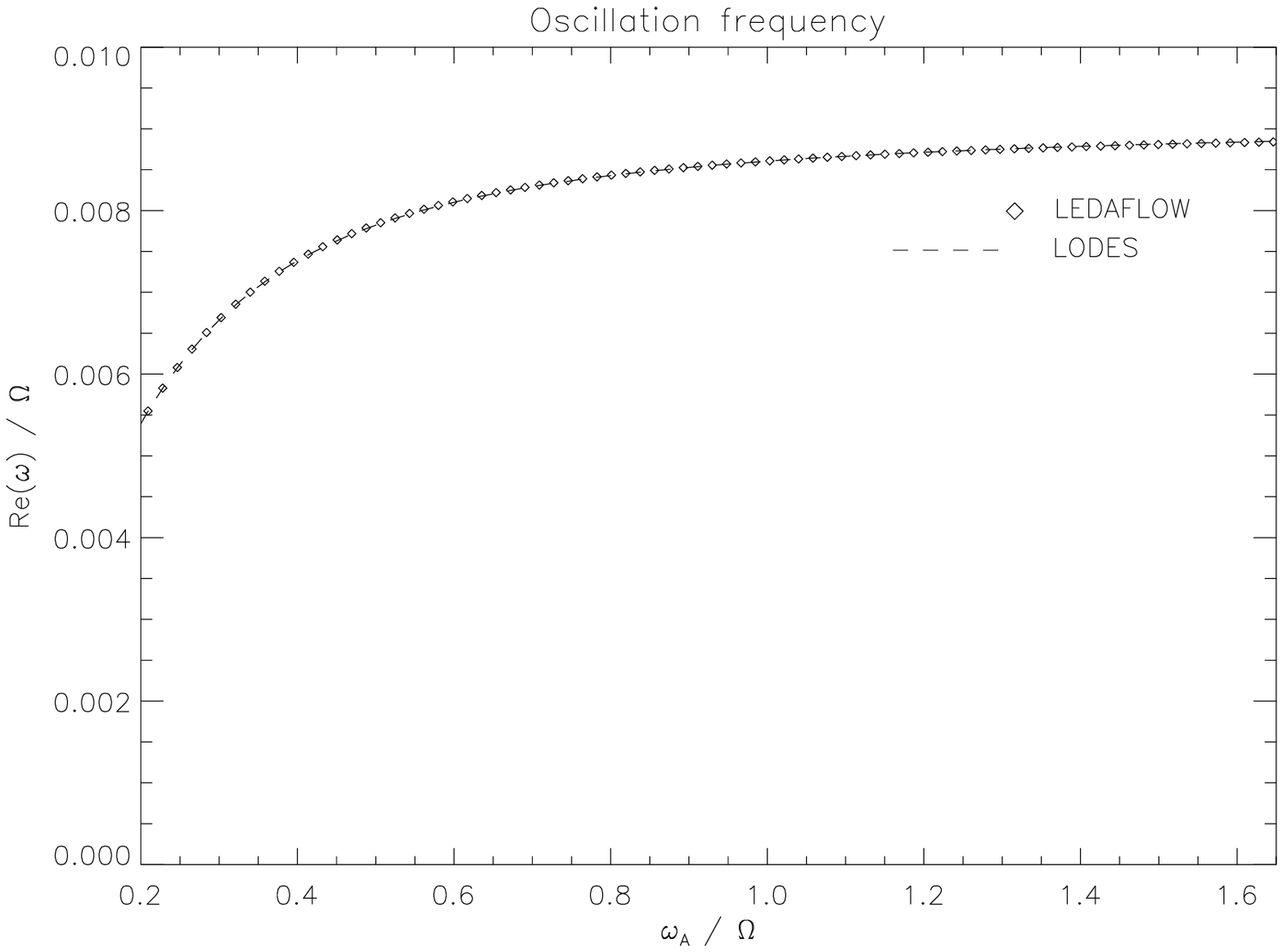, width=8cm}
  \end{tabular}
  \caption{The growth rate and the oscillation frequency of the most unstable mode 
           for disk model G, where the disk rotation is strong sub-Keplerian.}
  \label{fig:Vg3}
\end{figure*}

The results are shown in Figs.~\ref{fig:Vg1}-\ref{fig:Vg3}. Again, the growth rate 
(left panel) and oscillation frequency (right panel) are scaled with respect to the rotation 
frequency. These scaled quantities have been plotted as a function of the 
ratio of the Alfv\'en frequency to the rotation frequency.

Again, there is a perfect match (less than~1\% discrepancy) for the growth rate obtained with 
the three methods: LEDAFLOW, LODES and the $4^{\rm th}$ order polynomial~\eqref{eq:polynomialBH}.
Notice that the scaled growth rates in Figs.~\ref{fig:Vg1}-\ref{fig:Vg3} 
coincide although the deviation from a Keplerian disk is significant in cases F and G.
This can be explained by investigating the solution of the $4^{\mathrm{th}}$ order polynomial 
in the case of a weakly magnetized disk. The solution, scaled with respect to the rotation 
frequency, reads:
\begin{align}
   \frac{\omt^{2}}{\Om{}^{2}} = &
   \frac{\om{A}^{2}}{\Om{}^{2}} + \frac{k^{2}}{2(q^{2}+k^{2})} \Biggl\{
   \frac{\kappa^{2}}{\Om{}^{2}} + 
   \frac{1}{\Om{}^{2}}\left( \frac{\rho'}{\rho}V_{g} - \frac{\rho V_{g}^{2}}{\gamma p} \right) \nonumber \\
   & \\
   & \pm
   \sqrt{ \left[ \frac{\kappa^{2}}{\Om{}^{2}} + 
                 \frac{1}{\Om{}^{2}}\left( \frac{\rho'}{\rho}V_{g} - \frac{\rho V_{g}^{2}}{\gamma p} \right)
          \right]^{2}
          + 16 \frac{q^{2}+k^{2}}{k^{2}} \frac{\om{A}^{2}}{\Om{}^{2}} } \Biggr\}. \nonumber
   \label{eq:polynomialsolution}
\end{align}
For the chosen power-law of the toroidal velocity, the epicyclic frequency~$\kappa^{2}=\Om{}^{2}$. 
In the case of a weakly magnetized disk and a constant entropy, the term~$\rho' V_{g}/\rho - \rho V_{g}^{2}/(\gamma p)$ 
can be neglected. Furthermore, notice that for a pure hydrodynamical case this term is the Brunt-V\"aisala frequency.
For the calculations in this paper, this frequency is equal to zero because we consider disks with constant
entropy.

This results in a equation that only depends on the ratio $\om{A}/\Om{}$. Remember that the domain of this ratio 
has been kept the same in calculations~E, F and G.

The oscillation frequencies in Figs.~\ref{fig:Vg1}-\ref{fig:Vg3} obtained from LEDAFLOW and LODES match 
perfectly (less than 1\%). These figures show an overall increase in the value of the scaled oscillation frequency 
as one increases the value of $\epsilon$. Hence, the overstable nature of the MRI is even more significant in 
sub-Keplerian disks with a dominant toroidal magnetic field component.

\section{Conclusions}
We have considered the magnetorotational overstability in magnetized accretion disks
with a toroidal magnetic field component. We have used results from the spectral code LEDAFLOW
and compared them with more approximate solutions of both the 
fourth and sixth order polynomial local dispersion equation. In our calculations,
we have considered axisymmetric perturbations $(m=0)$ within a disk in the 
cylindrical limit. The most important results are summarised below:

\begin{itemize}
\item Magnetorotational instabilities are present in both sub-Keplerian and Keplerian
      rotating accretion disks.
\item The magnetorotational instability is also present in Keplerian rotating disks close
      to equipartition, when the toroidal magnetic field strength dominates the total 
      magnetic field strength.
\item The oscillation frequency of the instability is nonzero due to the presence of a 
      toroidal magnetic field component for both sub-Keplerian and Keplerian rotating 
      disks.
\end{itemize}

In an accompanying paper by van der Swaluw et al. (\cite{SBK}), we will consider
accretion disk models that also allow for convective instabilities, and analyse their
relation with respect to MRI modes. All our solvers can be used to 
study non-axisymmetric perturbations $(m\ne 0)$. This is left to future work.

\begin{acknowledgements}
J.W.S. Blokland, R. Keppens and J.P. Goedbloed carried out this work within the framework of the
European Fusion Programme, is supported by the European Communities under the contract of the Association 
between EURATOM/FOM. Views and opinions expressed herein do not necessarily reflect those of the European
Commission. E. van der Swaluw did this research in the FOM projectruimte on `Magnetoseismology of
accretion disks', a collaborative project between R. Keppens (FOM Institute Rijnhuizen, Nieuwegein) and 
N. Langer (Astronomical Institute Utrecht). This work is part of the research programme of the `Stichting voor
Fundamenteel Onderzoek der Materie (FOM)', which is financially supported by the `Nederlandse Organisatie voor 
Wetenschappelijk Onderzoek (NWO)'.
\end{acknowledgements}

\appendix
\section{Radial `wavenumber' \label{sec:radialwavenumber}}
To justify how we calculate the radial `wavenumber'~$q$ as discussed in subsection~\ref{subsec:coupling}, 
we discuss the WKB approximation in detail. Instead of using the system of first order differential 
equations~\eqref{eq:fod} we use the equivalent generalized Hain-L\"ust equation,
\begin{equation}
   (f \chi')' + \hat{g}\chi = 0,
   \label{eq:hl}
\end{equation}
where
\begin{align}
   f       & \equiv \frac{AS}{rD}, \\
   \hat{g} & \equiv \frac{-r}{ASD} \left[ C^{2} + DE - \frac{ASD}{r} \left( \frac{C}{D} \right)' \right].
\end{align}
This equivalent equation is preferred over system~\eqref{eq:fod} for a more straightforward WKB analysis.
In standard WKB analysis for the generalized Hain-L\"ust equation~\eqref{eq:hl} we write the solution 
in the following form:
\begin{equation}
   \chi (r) = p(r) \exp \left[ \mathrm{i} \int_{r_{0}}^{r} q(r) \mathrm{d}r \right],
\end{equation}
where~$r_{0}$ is the radial location of the inner boundary. Inserting this solution in the 
generalized Hain-L\"ust equation~\eqref{eq:hl} gives
\begin{equation}
   -fq^{2}p + \hat{g}p + fp'' + f'p'  + \mathrm{i} \left( 2fqp' + f'qp + fq'p \right) = 0.
\end{equation}
Introducing a length scale $L$ for the radial background variation, we can neglect the second order
terms~$fp'' + f'p'$ compared to the first order ones if $qL \gg 1$. This kind of ordering yields the
following expressions for the radial `wavenumber'~$q$ and the amplitude~$p$:
\begin{equation}
  q \approx \pm \sqrt{\frac{\hat{g}}{f}} \quad \text{and} \quad p \approx \frac{1}{ (f\hat{g})^{1/4} }.
\end{equation}

\begin{figure}
   \centering
   \epsfig{file=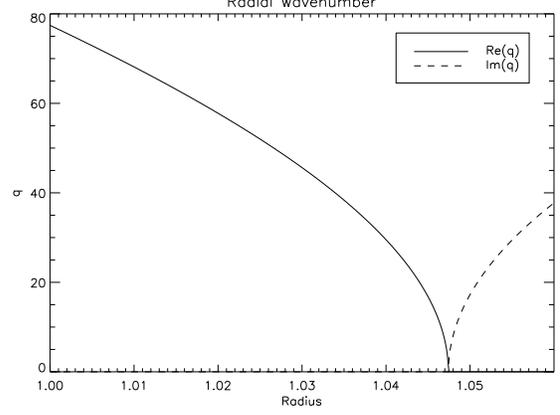, width=8cm}
   \caption{The radial `wavenumber'~$q$ of the eigenfunction~$\chi$ shown in Fig.~\ref{fig:mrichi}.
            The growth rate is $0.728\Omega$ .}
   \label{fig:mriq}
\end{figure}

Using the expression for $q$, we can plot this quantity for a given radial domain. This is shown in
Fig.~\ref{fig:mriq} for the eigenfunction~$\chi$ plotted in Fig.~\ref{fig:mrichi}.
Here, we applied the same radial domain as used in the LEDAFLOW calculation. The plot shows that
at a certain radius~$r_{1}$ the radial `wavenumber' becomes zero. Close to this radius the WKB 
approximation fails, because $qL \gg 1$ is no longer satisfied. However, we can find
an analytical solution close to this radius. To find the analytical solution we write the 
generalized Hain-L\"ust equation~\eqref{eq:hl} in the following form
\begin{equation}
   \frac{\mathrm{d}^{2}\chi}{\mathrm{d}R^{2}} + G \chi = 0,
   \label{eq:hlmod}
\end{equation}
where
\begin{equation}
   R' \equiv \frac{1}{f}, \quad \text{and} \quad
   G  \equiv f\hat{g}.
\end{equation}
Both the radial `wavenumber'~$q$ and the function~$G$ become zero at the radius~$r_{1}$ as expected.
We apply a Taylor expansion of $G$ about $R_{1} ( \equiv R(r_{1}))$ and insert this in the generalized
Hain-L\"ust equation~\eqref{eq:hlmod}. This yields
\begin{equation}
   \frac{\mathrm{d}^{2}\chi}{\mathrm{d}z^{2}} - z \chi = 0,
   \label{eq:hlmodR1}
\end{equation}
where
\begin{equation}
   z \equiv \left( \left| \frac{\mathrm{d}G}{\mathrm{d}R} \right|_{R=R_{1}} \right)^{1/3} \left( R - R_{1} \right).
\end{equation}
The analytical solution of this differential equation is the Airy function,
\begin{equation}
   Ai(z) = \frac{1}{\pi} \int_{0}^{\infty} \cos \left( \tfrac{1}{3}s^{3} + sz \right) \mathrm{d}s,
\end{equation}
which for large~$|z|$ has the asymptotic form
\begin{equation}
\begin{aligned}
   Ai(z) & \sim \frac{1}{2\sqrt{\pi}  z^{1/4} } \exp \left[ -\frac{2}{3}  z^{3/2}                    \right] & \text{for } z & > 0, \\
   Ai(z) & \sim \frac{1}{ \sqrt{\pi}(-z)^{1/4}} \sin \left(  \frac{2}{3}(-z)^{3/2} + \frac{1}{4} \pi \right) & \text{for } z & < 0.
\end{aligned}
\end{equation}
We now calculate the integral
\begin{equation}
   \int_{r}^{r_{1}} q(r) \mathrm{d}r = \frac{2}{3} (-z)^{3/2},
\end{equation}
which, for $R<0$, results in the solution
\begin{equation}
   \chi (r) = \frac{A_{0}}{ (f\hat{g})^{1/4} } \sin \left( \int_{r}^{r_{1}} q(r)\mathrm{d}r + \frac{1}{4}\pi \right),
\end{equation}
where $A_{0}$ is constant. We apply to this function the same boundary condition as in LEDAFLOW at the inner boundary. 
This means that~$\chi(r)$ is zero at the inner boundary, which yields
\begin{equation}
   \int_{r_{0}}^{r_{1}} q(r)\mathrm{d}r = -\frac{1}{4}\pi + n\pi,
\end{equation}
where~$n$ is a natural number. If we numerically calculate this integral for the radial `wavenumber'~$q$ 
plotted in Fig.~\ref{fig:mriq}, we find that it is approximately 0.76$\pi$.
For numerical integration we used Simpson's rule (Press et al. \cite{PTVF}). Thus determining the radial `wavenumber' 
as discussed in subsection~\ref{subsec:coupling} gives approximately the right value for~$q$. Hence, even for this most 
unstable mode, the local WKB analysis is in excellent agreement with the numerical solution.

\end{document}

%% file: spectrum.pstex_t
\begin{picture}(0,0)%
\includegraphics{spectrum.pstex}%
\end{picture}%
\setlength{\unitlength}{1973sp}%
\begingroup\makeatletter\ifx\SetFigFont\undefined%
\gdef\SetFigFont#1#2#3#4#5{%
  \reset@font\fontsize{#1}{#2pt}%
  \fontfamily{#3}\fontseries{#4}\fontshape{#5}%
  \selectfont}%
\fi\endgroup%
\begin{picture}(7074,520)(439,106)
\put(2626,164){\makebox(0,0)[lb]{\smash{{\SetFigFont{6}{7.2}{\familydefault}{\mddefault}{\updefault}{$\Om{S}^{-}$}}}}}
\put(4876,164){\makebox(0,0)[lb]{\smash{{\SetFigFont{6}{7.2}{\familydefault}{\mddefault}{\updefault}{$\Om{S}^{+}$}}}}}
\put(6151,164){\makebox(0,0)[lb]{\smash{{\SetFigFont{6}{7.2}{\familydefault}{\mddefault}{\updefault}{$\Om{A}^{+}$}}}}}
\put(7126,239){\makebox(0,0)[lb]{\smash{{\SetFigFont{6}{7.2}{\rmdefault}{\mddefault}{\updefault}{$\omega$}}}}}
\put(3826,164){\makebox(0,0)[lb]{\smash{{\SetFigFont{6}{7.2}{\familydefault}{\mddefault}{\updefault}{$\Om{0}$}}}}}
\put(1426,164){\makebox(0,0)[lb]{\smash{{\SetFigFont{6}{7.2}{\familydefault}{\mddefault}{\updefault}{$\Om{A}^{-}$}}}}}
\end{picture}%